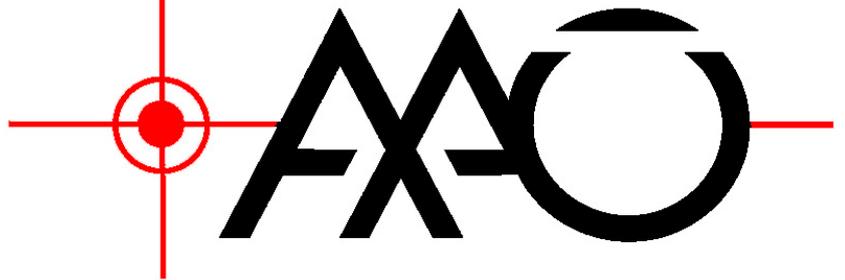

# AAO NEWSLETTER

**ANGLO-AUSTRALIAN OBSERVATORY**

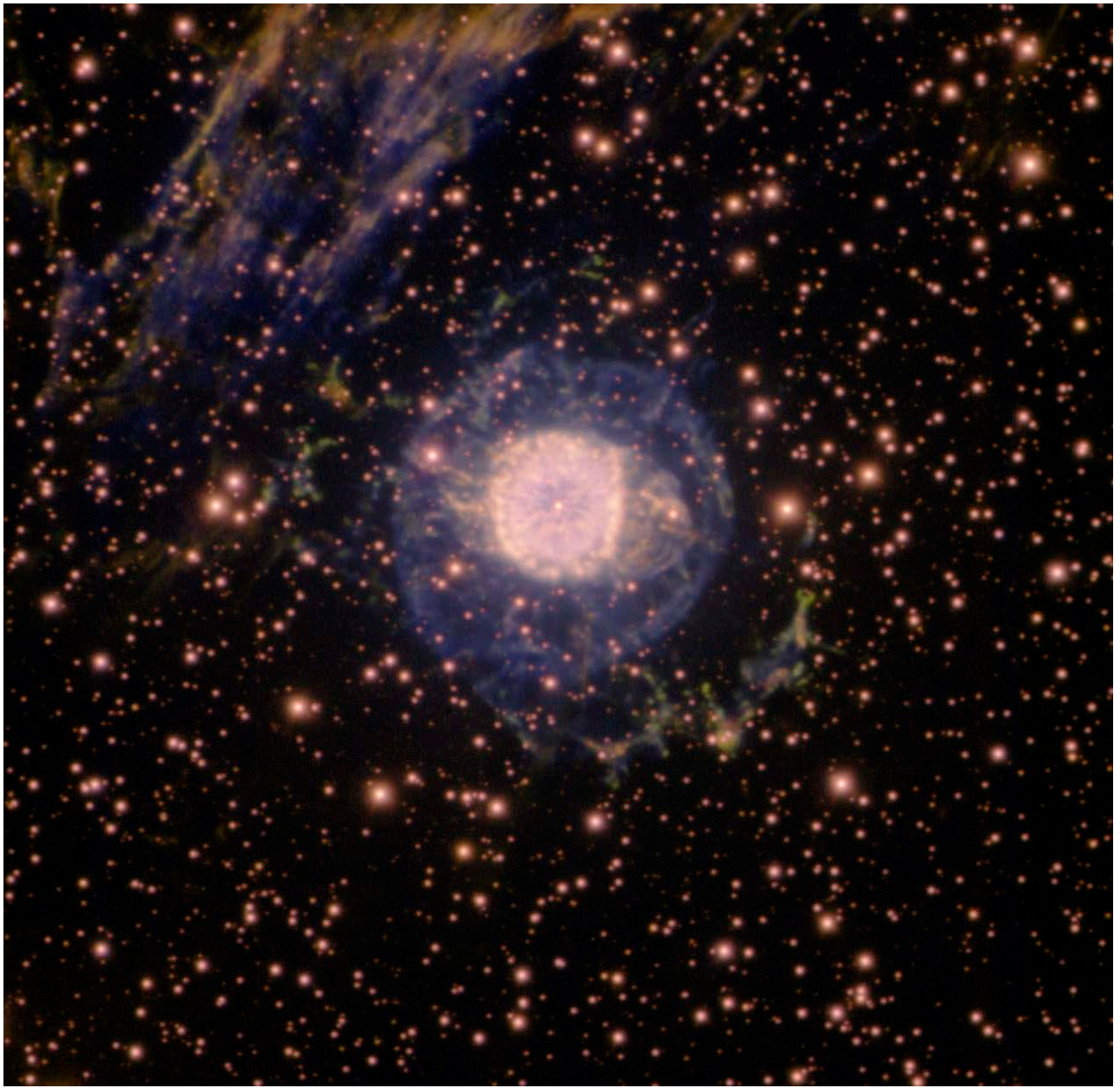

Gemini South GMOS image of NGC 6751, the "Glowing Eye" planetary nebula proposed by the winner of the Australian Gemini Office's IYA contest, Daniel Tran. This is a composite of images taken in three optical narrow-band filters, which isolated the Hα (yellow), [S II] (red), and [O III] (blue) emission lines. Image Credit: Daniel Tran (PAL College), Travis Rector (U. Alaska Anchorage), and Terry Bridges (Queen's U.).

# DIRECTOR'S MESSAGE

This is the last newsletter for the Anglo-Australian Observatory. On the 30 June the 36-year association between the United Kingdom and Australia on the Anglo-Australian Telescope. From 1 July the AAO becomes the Australian Astronomical Observatory, and the AAT becomes a wholly Australian facility.

The Anglo-Australian connection has been remarkably successful and productive over these three and a half decades. In order to celebrate this scientific partnership more fully, the AAO is holding a week-long symposium, "Celebrating the AAO: past, present and future", at Coonabarabran on 21-25 June. Everyone with any association with the AAO is very welcome to participate in this meetiing, which will review the history and accomplishments of the observatory, present some of the current research from the facilities both operated (AAT, UKST) and supported (Gemini, Magellan) by the AAO, and look to the future of the AAO, considering the development of the organization, new instruments for the various telescopes, and some of the ambitious new science goals. There will also be a variety of public talks and other events reflecting the AAO's commitment to public outreach and education. If you are interested in attending the symposium, either for the whole week or for just part, you can register on the web at http://www.aao.gov.au/AAO/AAO2010 - but note that places are limited, so sign up soon for this unique event!

The AAO is continuing to upgrade it's facilities with new instrumentation. HERMES, the major new instrument for the AAT, underwent its Preliminary Design Review at the end of February. Thanks to additional funding from Astronomy Australia Limited, provided under the EIF program, HERMES will now be a four-channel spectrograph, covering the range from 350nm to 950nm with four 20-30nm ranges. HERMES is expected to see first light in 2012.

In the meantime we have the CYCLOPS fibre feed for UCLES coming into service in semester 2010A, and the GNOSIS OH-suppression fibre feed for IRIS2 to come on-line in 2011. The AAO has also obtained support from AAL (again under EIF) to upgrade the detectors on AAOmega, increasing the effective efficiency by a factor of 2-3 longward of 850nm in the red channel and increasing the wavelength range to 1100nm. These new instruments and upgrades will keep the AAT highly competitive for years to come.

The AAO is also aiming to providing better research support for its users and for students. We currently have 3 UK and 3 Australian undergraduate summer scholars working at the AAO each year, a long-running tradition that has kicked-off the a number of successful careers (including that of the current Director!). This tradition will continue with the new AAO, supporting both Australian and international undergraduates. The AAO also provides two or three PhD top-up scholarships each year to encourage graduate students (and their supervisors) to work with AAO astronomers, and a Honours scholarship in conjunction with Macquarie University. A new program starting 2010 is the AAO Distinguished Visitors program. Each year, the AAO will support a small number of eminent researchers to make extended visits to the AAO. In 2010 we are fortunate to have Andrew Connolly, Richard Ellis, David Koo and Kim Vy Tran coming to work with astronomers at the AAO for periods of weeks to months.

Another AAO astronomer was also distinguished further. In the Australia Day (26 January) Honours list, Fred Watson was made a Member of the Order of Australia for his services to astronomy and public education. Congratulations to Fred Watson AM!

Matthew Colless

# UCLES OBSERVATIONS OF HELIUM-RICH SUBLUMINOUS B STARS

Simon Jeffery, Naslim Neelanamkodan,
Amir Ahmad, Xianfei Zhang
(Armargh Observatory)

**Normal subluminous B stars**

In recent years, subluminous B (sdB) stars have attracted attention from galactic and extra-galactic astronomers alike. Indeed, they pres-ented a puzzle when first discovered as faint blue stars in the galactic halo (Humason & Zwicky 1947). With luminosities between those of B-type main-sequence stars and white dwarfs, it took thirty years to demonstrate that they are extreme horizontal-branch stars, or stars of half a solar mass lying virtually on the helium main-sequence (Heber 2009).

The first problem is how such a star loses its hydrogen envelope following core-helium ignition. Whatever the process, it cannot be rare; sdB stars outnumber QSOs and white dwarfs in the Palomar-Green and other surveys at least to Mb<18 (Green et al. 1986, Beers et al. 1992). Meanwhile, giant elliptical galaxies offer a quite different puzzle -- the existence of a significant ultraviolet contribution to their integrated light, completely at variance with the output of the old cold stars which dominate at optical wavelengths (O'Connell 1999).

Over the last decade, renewed interest has come from a series of discoveries:
1) the majority of sdB stars are formed in binaries of at least four different types (Han et al. 2002).
2) a significant fraction of sdB stars pulsate, allowing high precision measurements to be made using the techniques of asteroseismology (Kilkenny et al. 1997, Brassard et al. 2001).
3) the elliptical galaxy UV-excess problem can be successfully explained if a substantial population of sdB stars is always present in old stellar environments (Yi et al. 1997, Brown et al. 2003).
4) stellar population calculations which include binary star evolution give good agreement with the numbers and types of sdB stars observed in our own Galaxy and required by the UV-excess observations (Han et al. 2002, 2003, 2007).

**Helium-rich subluminous B stars**

So what else is left to learn? Although the sdB stars are nearly naked helium stars, the majority retain a thin veneer of hydrogen (<0.01 $M_\odot$). With radiative envelopes (no convection) and a high surface gravity, whatever helium there is sinks below the surface and makes the majority quite deficient in helium. Diffusion theory which includes radiative forces acting on ions explains this well (Heber 1986). So the existence within the PG and other surveys of a fraction (~10%, Green et al. 1986) of "helium-rich" sdB and sdO stars poses a major challenge (Fig. 1). Some of these stars are so helium-rich that hydrogen Balmer lines, normally dominant in B-type spectra, are undetectable (Fig. 2).

Might these stars be ultra-extreme sdB stars in which the hydrogen has been completely lost? Or are they evolving towards the sdB domain, so that the helium has not had time to sink out of sight? Or will they by-pass the sdB domain completely and become something else? To address these questions, we needed better information about chemical abundances, temperatures, surface gravities, and statistics. While establishing some basic properties from low resolution data, we encountered a surprise and a puzzle. The surprise was that the prototype He-sdB, PG1544+488, is a binary containing not one but two hot subdwarfs, both extremely helium rich. With a mass ratio close to unity and a period of 12 hours (Ahmad et al. 2004), stellar evolution theory has still to explain how to make two low-mass helium stars at exactly the same time. So far, PG1544+488 remains unique; the majority of HesdBs appear to be single stars.

The puzzle was that, for He-sdBs in general, the effective temperatures we obtained from optical data (Ahmad et al. 2003) were substantially lower than those obtained for three He-sdB stars observed with FUSE (Lanz et al. 2004). This could be explained by the choice of model atmosphere; whenever hydrogen is absent, the structure of a stellar atmosphere becomes critically sensitive to the abundance of other high-opacity ions, and strongly affects the emergent spectrum.

**Flashers or mergers?**

With these first results in place, our first conclusions were that He-sdB stars have lower surface gravities than classical sdB stars and that they should be considered alongside the hotter helium-rich sdO (He-sdO) stars. Originally, it had been thought that the latter represent stars evolving away from the sdB domain to become white dwarfs. However, the He-sdB and He-sdO stars seem to form a continuous sequence in both spectral type and in the temperature-gravity diagram (Fig. 1).

Several models of stellar evolution might explain such a distribution in terms of contraction towards and onto the helium main sequence.

One model (Lanz et al. 2004) considers the evolution of a red giant core which has been stripped of its hydrogen envelope. If helium ignites (off center) after the star has started contracting to become a helium white dwarf, this "late flasher" will initially expand to become a yellow giant. It then contracts toward the helium main sequence as the helium-burning shell flames inwards to the stellar core. All such stars should lie in a very narrow mass range around 0.48 $M_\odot$. During the yellow giant phase, convection should force nitrogen-rich helium to the surface and ingest any residual hydrogen. A small amount of carbon is produced in the helium shell flash, which convection might transport to the surface. For this model to work, the red giant progenitor requires a hydrogen-stripping agent, which could be a low-mass companion or enhanced mass loss from rapid rotation (Miller Bertolami et al. 2008).

Another model considers what happens when two helium white dwarfs merge (Iben 1990, Saio & Jeffery 2000). In such a merged star, helium again ignites off center and produces a similar evolution (Fig. 1). The mass range should be larger; the surface should be nitrogen rich, possibly with some hydrogen. A small amount of carbon is produced in the first helium shell flash; the models do not show significant surface enrichment. These stars will have a wider mass range (possibly 0.4 - 0.7 $M_\odot$).

A third and recent suggestion considers what would happen when a white dwarf with a low mass CO core merges with a helium white dwarf (Justham & Podsiadlowski 2010). Such a merger produces a qualitatively similar evolution and mass range as the HeWD merger. With significant carbon already present in one of the progenitor components, it may easier to deliver a larger quantity of carbon to the surface.

**Chemical signatures**

To address these questions, we obtained AAT+UCLES observations of several He-sdBs to measure more precisely their H, He, C, N and O abundances, along with an indication of initial metallicity. Some of these spectra are shown in Fig. 2. Note the variety in helium to hydrogen ratio. The range of carbon and nitrogen is less obvious. It is clear that helium subdwarfs (He-sd's) comprise an heterogenous group. Taken together with recent results for He-sdO stars (Stroeer et al. 2007), the group contains N-rich stars, C-rich stars, stars with moderate hydrogen depletion, and stars with extreme hydrogen depletion (Figs. 1 - 3). The most numerous He-sd's are both very helium-rich and carbon-rich, but significant numbers show no carbon enhancement and only some helium enhancement.

Nitrogen-rich surfaces are simply understood if the visible helium has been produced by the CNO-process (hydrogen burning); all original carbon and oxygen is converted to nitrogen. Carbon-rich surfaces are harder to explain since 3α processed material must be exposed at the surface, possibly by dredge-up after helium ignition but while the surface is still cool. The models seem very sensitive to the convection physics.

This variety of surface composition amongst stars which lie within a tightly constrained temperature-gravity locus could be produced by a single process in which the degree of core-envelope-surface mixing is indeterminate, as precipitated by a range of initial masses in a binary merger, for example. Alternatively, several evolution paths may converge, differentiated only by the final surface composition and, possibly, the mass of the product. From their distribution in gravity and temperature, the He-sd's appear to lie closer to the evolution track for the white dwarf merger than the late hot flasher (Naslim et al. 2010), although it is not clear what rôle is played by the choice of physics adopted in the respective simulations.

What is evident from these studies is that, far from evolving from the extreme horizontal branch to the white dwarf domain (i.e. post-classical sdB stars), the helium-rich subdwarfs represent a distinct group of stars evolving toward and onto the helium main sequence. Continuing observations with AAT+UCLES and other telescopes will explore the systematics of surface composition in more detail, and hence attempt to disentangle which evolutionary pathways produce these unusual stars.

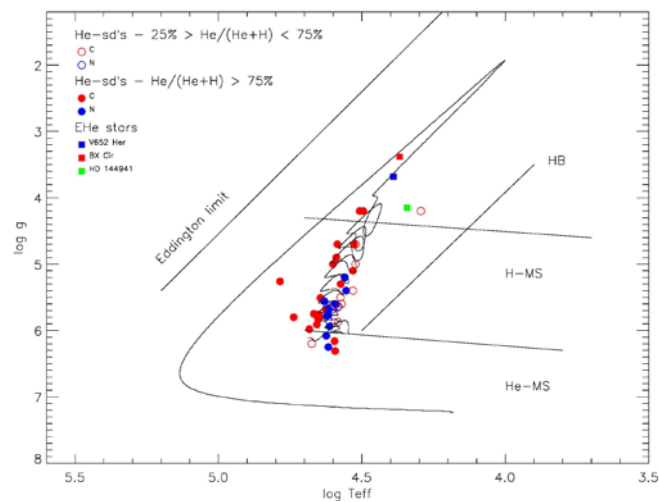

**Figure 1:** *Surface gravities and effective temperatures of helium-rich sdO (Stroeer et al. 2007), sdB (Ahmad et al. 2006) and related stars compared with the simulated evolution of a double helium white dwarf merger (Saio & Jeffery 2000).*

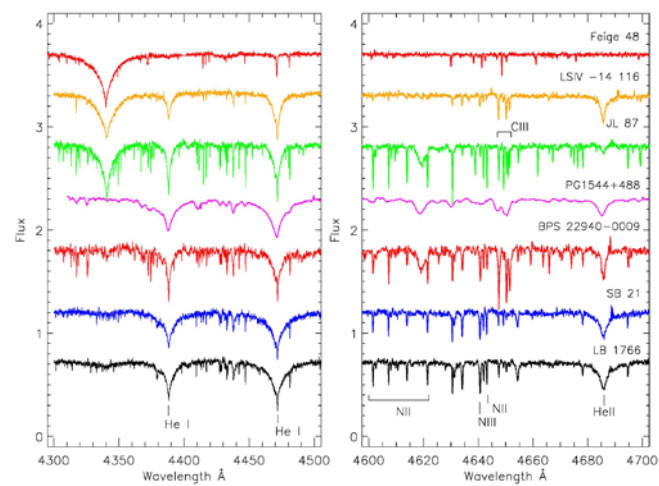

**Figure 2:** *High-resolution spectra of several helium-rich subdwarf B stars, obtained with AAT+UCLES and WHT+ISIS. Key absorption lines of H, HeI, HeII, C and N are marked. For comparison, the classical H-rich sdB star Feige 48 is also shown (top).*

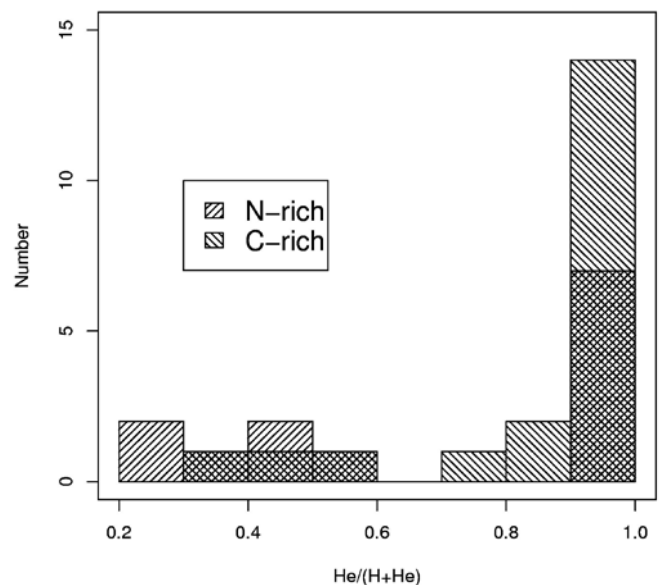

**Figure 3:** *Distribution of carbon- and nitrogen-rich He-subdwarfs as a function of surface helium abundance (Stroeer et al. 2007, Naslim et al. 2010)*

# DISCOVERY OF LARGE-SCALE GRAVITATIONAL INFALL IN A MASSIVE PROTO-STELLAR CLUSTER

Peter Barnes (U. Florida), Yoshi Yonekura (Ibaraki U.), Stuart Ryder, Andrew Hopkins (AAO), Yosuke Miyamoto, Naoko Furukawa, & Yasuo Fukui (Nagoya U.)

**Abstract**
From AAT and Mopra observations, we have identified a young massive star-forming cloud as undergoing a large-scale gravitational collapse, likely on the way to forming a massive young star cluster. Both the size scale and the mass infall rate may be new records among Galactic star-forming regions. This object promises to be an important testbed for refining theories of massive cluster formation.

**Massive Star Formation in the Milky Way**
Compared to our understanding of how Sun-like ("low-mass") stars form out of cold, molecular gas, the formation of massive stars and star clusters, which may dominate and drive the Galactic ecology with their high luminosities, massive winds, and chemically enriched ejecta, is not nearly so well understood. This enigma has 3 causes: the relative rarity of massive star formation, the rapidity of massive star evolution, and the confusing phenomenology of the formation process itself. Because of the first 2 reasons, the typical massive star formation site lies more than 10 times further away than many low-mass protostars, further limiting our ability to decipher the phenomena we see. Thus, there is little clear consensus on even the basic formation mechanism, whether through gravitationally-powered accretion disks (e.g. McKee & Tan 2003), competitive accretion of ambient cluster gas (Schmeja & Klessen 2004), or more radical theories. A vast range of parameters, such as formation timescale or accretion rate, are debated: e.g., is the overall timescale for star-cluster formation a few (Elmegreen 2007) or many (Tan, Krumholz & McKee 2006) free-fall times? The influence of feedback in setting both the stellar initial mass function (IMF), including its upper limit, and the efficiency of star formation in clusters, is uncertain, as is the universality or variability of the IMF (Hoversten & Glazebrook 2008).

**CHaMP**
To address these issues, we designed the Census of High- and Medium-mass Protostars as the largest, most uniform, and least biased survey of massive Galactic star-forming regions to date. We reasoned that only with an unbiased survey could we hope to construct a comprehensive paradigm for massive star- and cluster-formation, including the identification of all significant stages in massive star formation, and their lifetimes. CHaMP was based on the Nanten molecular cloud surveys (Yonekura et al 2005) of a large portion of the southern Milky Way in Carina, Vela, & Centaurus, completely covering a 20°x6° window that samples approximately 5% of the total star formation of our Galaxy (Fig. 1). Using the Nanten maps as 3′-resolution finder charts, we identified 209 massive, dense molecular clumps that must include all likely sites of massive star formation in this window. We then used the ATNF's Mopra antenna to zoom in, at 36″ resolution, to the 118 brightest of these.

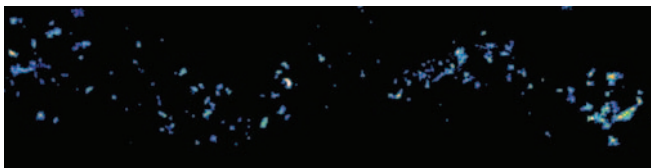

**Figure 1:** *Integrated intensity image of C18O over a 20x6 degree window towards the Carina arm, observed by the Nanten telescope. Note that the C18O emission only covers ~2% of the Galactic Plane, much less than 13CO or 12CO. With Nanten maps as "finder charts", CHaMP at Mopra has been able to efficiently map the dense gas tracers in >100 molecular clouds at high resolution and sensitivity.*

**BYF 73 and Infall Modelling**
Among these clumps, we noticed very unusual spectral line profiles of the HCO+ molecule in one source, the 73rd on our list. They showed the classic (Zhou et al 1994) inverse P Cygni profile seen in lower-mass protostars undergoing gravitational collapse, but on a parsec-wide scale and over a much larger velocity range than usually seen (Figs. 2, 3). When we modelled these line profiles with a radiative transfer code that allows for infall motions (De Vries & Myers 2005), we found most of the ~20,000 M¤ clump to have a high infall speed that implied a mass infall rate that was either a new record, or close to it: 3.4 x 10-2 M¤/yr. Moreover, gravitational infall in the gas seems to be the only option for BYF 73: it is far too massive to obtain sufficient support from any reasonable estimates of rotational, thermal, or magnetic field pressure.

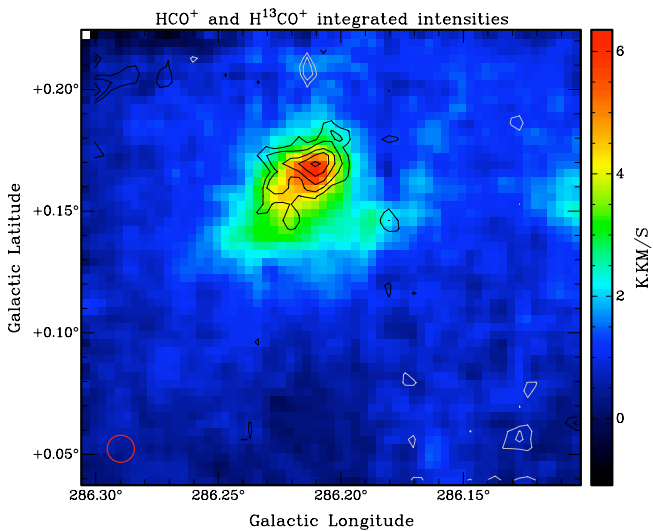

**Figure 2:** *Mopra HCO+J=1®0 integrated intensity from BYF 73 over an 11' field, on the T*A scale as given by the colour bar. The integration is over the range -23.20 to -16.63 km/s or 58 channels, yielding an rms noise level 0.16 K km/s: hence the widespread low-level emission above ~0.5 K km/s is real. Contours: Mopra HCO+ J=1®0 integrated intensity in T*A, levels are (grey) -0.5, -0.35, (black) 0.35, 0.5, 0.7, 0.9, and 1.1 K km/s. The integration is from -21.94 to -17.86 km/s, giving an rms noise level 0.12 K km/s. The smoothed Mopra HPBW for both datasets (40″) is shown for reference in the lower-left corner.*

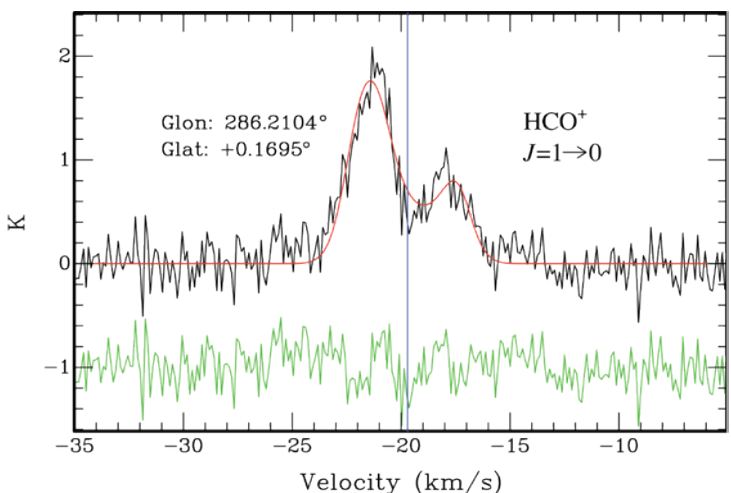

**Figure 3:** *Sample Mopra HCO+ J=1®0 spectrum of BYF 73 in black at the peak H13CO+ position. The radiative transfer model fit at this point (De Vries & Myers 2005) is shown in red, and the residual spectrum (data-model) is shown in green, offset 1K below the HCO+ spectrum. The vertical blue line indicates the systemic velocity at VLSR = -19.7km/s, also from the model fits.*

### IRIS2 Images

By itself, the molecular data and modelling would have been an interesting result. But the clincher came when we combined our Mopra data with narrowband 2µm imagery using the AAT's IRIS2 camera. Using only a few hours of Service Time to image BYF 73 and a few other interesting CHaMP sources, we obtained snapshots of Br-ā, H2 S(1) v=1®0 & v=2®1, and an equivalent narrowband image of line-free continuum near 2.3µm. When we aligned, calibrated, and subtracted the continuum from the line images, we obtained the remarkable result in Figure 4, where we have overlaid the Mopra HCO+ & H13CO+ contours with the 3-colour near-IR image. This combined image shows, in a very clean way, the formation of an HII region at the edge of a molecular cloud, surrounded by a cocoon of presumably shocked H2 ahead of the ionisation front, driven from the already-revealed massive young stars in the Br-ā nebula. An IRIS2 long-slit spectrum across these features (Fig. 5) reveals several more H2 lines, whose ratios indicate a temperature in the pre-ionised molecular gas that may exceed 4000 K. But most significant is the location of the centre of the molecular infall revealed in the Mopra maps: this is precisely where we have a very deeply embedded IR nebula, and stars with very unusual colours.

### Status of the Cloud

Indeed, at mid- and far-IR wavelengths we can see that this infall centre is the most luminous source of the whole cloud, and is extremely red even at mid-IR wavelengths. We calculate that the release of gravitational energy alone accounts for at least 4% of the total luminosity. If the star formation in BYF 73 turns out be as efficient as in other massive, dense molecular clouds, then we might expect ~6,000 M¤ of gas to be turned into stars. Even at a fraction of this efficiency, what we seem to have in BYF 73 is the precursor to a massive, rich, young stellar cluster, before nearly all of the usual hallmarks of such a cluster have had time to develop. However, the speed of the infall, ~1 km/s, is ~20x less than in a purely dynamical collapse, and so the timescale for cluster formation (i.e., until the infalling gas supply runs out) is quite long, ~0.6 Myr, compared to what is predicted by some "prompt" models of massive cluster formation. These results on BYF 73 have just appeared in MNRAS (Barnes et al 2010).

### Next Steps

During our IRIS2 Service Time, we also imaged a few other CHaMP clumps in the same way. Preliminary comparisons of these images with our Mopra data contains more surprises. We see several instances of cocooning, and sometimes filamentary, H2 emission surrounding other Br-ā nebulae, many of which are associated with bright HCO+ emission. At the same time, we see stand-alone H2 emission nebulae at positions away from HCO+ emission peaks. This suggests some systematic evolutionary effects, although we will need to perform a complete near-IR survey of our CHaMP clumps to more precisely characterise these interrelationships. We are planning to use IRIS2 in the near future to do this, and anticipate that the AAT will make a major contribution to CHaMP's goal of better understanding many aspects of massive star formation.

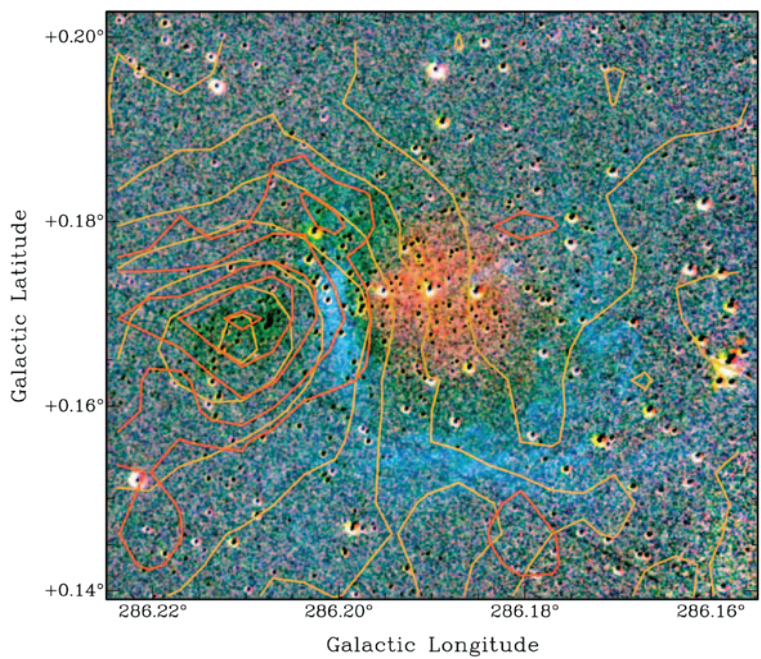

**Figure 4:** *RGB-pseudocolour image of BYF 73 in K-band spectral lines. Here Br-g is shown as red, and H2 S(1) is shown as green (v=1®0) & blue (v=2®1). Contours are overlaid from Mopra HCO+ (gold) and H13CO+ (red) integrated intensities (levels as in Fig. 2).*

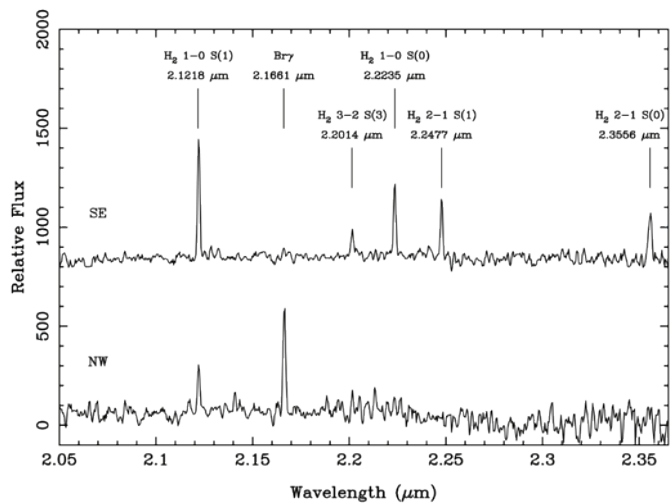

**Figure 5:** *Sample spectra from an IRIS2 long-slit spectral image, on a relative flux scale to indicate line ratios. The upper spectrum (labelled "SE") is close to the bright H2 interface between the HII region and the cold molecular cloud, while the lower spectrum (labelled "NW") is near the peak of the HII region.*

### HERMES: a multi-object High-resolution spectrograph for the AAT.

Ken Freeman (ANU) & Joss Bland-Hawthorn (University of Sydney), HERMES Project Scientists.
Sam Barden (AAO), HERMES Technical Team Leader

HERMES (High Efficiency and Resolution Multi-Element Spectrograph) is a new facility-class instrument currently under development at the AAO. HERMES uses the existing 2dF fiber positioner and a new spectrometer which allows the simultaneous spectroscopic observation of nearly 400 targets in 4 separate simultaneous wavelength channels at a spectral resolving power of about 30,000. The dispersing elements are VPH gratings, and the four wavelength channels are separated by dichroic beamsplitters. Figure 1 shows the optical layout. The target for HERMES performance is a signal-to-noise ratio of 100 per resolution element for a 60-minute integration of a star with V = 14.

The HERMES project has been ongoing since early 2008. A Preliminary Design Review will take place in Feb 2010. Commissioning of the instrument is expected to occur in 2012 with the start of science operations in the later part of 2012.

The primary science driver for the instrument is to reconstruct the chemical, dynamical and star formation history of the Milky Way through a major survey of about a million stars (the Galactic Archaeology or GA survey) for chemical tagging and velocity measurement. We can use chemical tagging techniques (Freeman & Bland-Hawthorn 2002, Bland-Hawthorn & Freeman 2004) to identify families of stars which were born together and have subsequently drifted apart. Stars which were born together have very similar chemical compositions (eg De Silva et al 2009). The chemical signature of such a set of stars is defined by several groups of chemical elements which do not vary in lockstep from star to star. We can measure abundances for about 25 elements: the "chemical space" which these elements define has about 8 independent dimensions. Together with the usual positional and kinematical phase space, each star will then be identified by its position in a 14-dimensional space which we can use to associate it with its siblings. Debris of tidally disrupted dwarf galaxies and globular clusters will also be recognizable by the loci of stars in this multi-dimensional space (eg Venn et al 2004, Wylie-de Boer et al 2010).

The four wavelength channels cover a total of about 100 nm. The wavelength channels for the GA survey are given in Table 1, and were chosen to include lines of all major element groups (light elements like Li, Na and Al; alpha-elements including O, Mg, Si, Ca; iron-peak elements, light and heavy s-process elements, r-process elements). The vast volume of stellar abundance data from the GA survey is likely to be used for a wide range of stellar astrophysics. In addition to the observations for the GA study, HERMES is ideal for many other high resolution multi-object spectroscopic studies. The individual channels are therefore designed to handle much wider wavelength bands in order to enable scientific programs which may need other wavelength bands and other VPH gratings and beamsplitters.

**Table 1: HERMES Wavelength Channels**

| Channel | Blue limit (nm) | Red limit (nm) |
|---|---|---|
| Blue | 468 | 487 |
| Green | 565 | 588 |
| Red | 648 | 674 |
| IR | 759 | 789 |

HERMES has an exciting synergy with the European GAIA astrometric mission which will deliver astrometry for about a billion stars in the next decade. For stars as bright as the HERMES GA survey stars (V < 14), GAIA's astrometry will be exceptionally precise, leading to distances with 1% errors and transverse velocities with errors < 1 km s -1. The accurate GAIA distances, combined with the element abundances and stellar parameters from HERMES, will allow derivation of accurate isochrone ages for a large fraction of the survey stars. Conversely, GAIA will detect a lot of phase space substructure among Galactic stars, and the element abundances from HERMES will show whether these substructures are remnants of tidally disrupted systems or come from dynamical resonances in the Galaxy. The accurate distances, ages, kinematics and abundances will also allow us to make a direct observational evaluation of the extent of radial mixing of stars in the Galactic disk (Sellwood & Binney 2002, Bland-Hawthorn et al 2010); this mechanism is potentially very significant for understanding the chemical evolution of the disk (eg Schönrich & Binney 2009).

HERMES will achieve much of the Galactic Archaeology science originally envisioned for the WFMOS project (Gemini and Subaru Observatories). WFMOS was a major item in the Decadal Plan for Australian astronomy, but is no longer viable for the near term. WFMOS included a high resolution spectrometer for a Galactic Archaeology survey. In terms of its simultaneous wavelength coverage and number of fibers, HERMES is a unique instrument for a large high resolution stellar survey. With its novel design and high throughput, HERMES will in many ways exceed the capabilities proposed for the WFMOS high resolution instrument, and is attracting a great deal of international interest.

In addition to the GA science and its immediate stellar physics applications, HERMES will be used for a variety of other science, including testing of nucleosynthesis theory, abundances of stars in the Magellanic Clouds, very metal-poor stars, stellar Doppler imaging, young associations, star clusters, properties of nearby dwarf galaxies, and studies of planetary nebulae and physics of the interstellar medium. A significant scientific infrastructure already exists within the Australian community preparing for HERMES-related science. Two dedicated postdoctoral staff are currently working on HERMES science and proposals have been submitted for several more at a number of Australian universities to complement and support the active and enthusiastic involvement of a core group of senior people in those universities. We expect to hold a HERMES science workshop in the first half of 2010. If observations with HERMES can contribute to your research, please come along to this workshop and talk about your ideas.

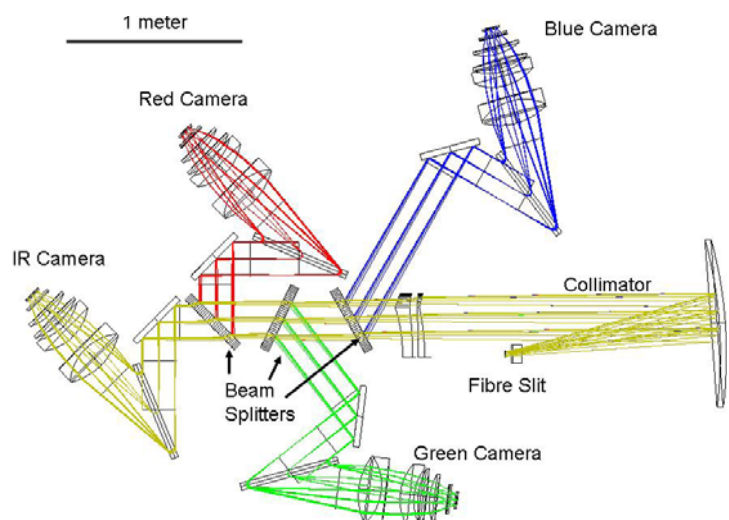

**Figure 1:** *schematic layout of the four-band HERMES spectrometer, showing the three dichroic beamsplitters, four VPH gratings and four cameras.*

# STARBUGS: FUTURE FIBRE POSITIONING TECHNOLOGY

Michael Goodwin, Jeroen Heijmans, Jurek Brzeski, Ian Saunders, Will Saunders (AAO)

Mentioning bugs and technology in the same sentence is generally not a good idea as you would find many organisations doing the best to keep bugs out of their technology. However, at the Anglo-Australian Observatory (AAO) we believe that infecting future optical telescopes with armies of self-motile miniature mechanical 'lift-and-step' bugs could bring substantial gains to astronomy.

These bugs, affectionately known as Starbugs, are an ongoing R&D effort at the AAO to deliver world class astronomical fibre positioning technology. The concept uses miniature bugs that scuttle around the telescope's focal plane catching the light from distant stars and galaxies to feed bundles of optical fibres. The light guided by the optical fibres is then fed into spectrographs for measurement and scientific analysis. The collecting end of each fibre has to be placed at precisely the right spot on the focal plane where each star or galaxy is expected to fall. Hence, for each different field of stars and galaxies, all the fibres, typically several hundred, need to be re-positioned appropriately.

The Anglo-Australian Telescope (AAT) is a leader in current fibre positioning technology for astronomy. The AAT uses fibres at prime focus that are re-positioned one-by-one using the "pick and place" mechanical arm of the 2 Degree Field (2dF) robotic fibre positioner. The 2dF currently feeds the AAOmega dual beam (blue and red arm) spectrograph with 392 fibres. The process is complex, time consuming and reduces the amount of observing time available to the astronomer (for the science case of many observation fields with typical exposures less than an hour). To drastically speed up the process making telescopes with fibre fed spectrographs significantly more productive, we are developing new miniature robotic prototypes to enable the simultaneous positioning of all fibres.

An example of the latest R&D efforts from the Starbugs Team is that developed during the design study proposal for the MANIFEST (many instrument fibre system) for the Giant Magellan Telescope (GMT). The prototype design is based on the 'hanging Starbug' and the 'lift-and-step' concepts, whereby the magnetic bugs hang inverted below a thin glass plate (see Figure 1). The prototypes are smaller that an 'AA' sized battery and therefore sufficiently compact to deploy on the GMT or other telescopes. The prototypes are constructed with piezoelectric ceramic cylinders that change shape when an electric field is applied. The careful manipulation of the electric field using voltage waveforms applied to electrodes on the outer and inner cylinders makes the robot perform microscopic 'lift-and-step' manoeuvres (see Figure 2).

The metrology and control of the Starbug prototypes in the laboratory are performed with developmental in-house software, see Figure 3. We have so far demonstrated a positional accuracy better than 1/10 of a pixel (about 10 microns on the field plate). The camera with wide-angle lens introduces aberrations requiring unique calibrations parameters that are determined using the Camera Calibration Toolbox for Matlab (Jean-Yves Bouguet). The software will also be used for autonomous testing with logged data processed and analysed offline in Matlab. Further software developments to be implemented include close-loop control, multiplexing and optimal routing algorithms. The metrology and algorithm development aspect is ongoing research to provide a generic platform to characterise developed prototypes.

In conclusion, one of the key advantages of the 'hanging Starbug' concept is that there are no retractors and no fibres crossing the focal plane. Thus very large numbers of fibre positions can be deployed. The Starbugs also have the advantage of significantly reducing the instrument weight (no need for a pick-and-place robot) as well as working in confined spaces. Each Starbug would have a 'patrol area' to simplify and speed up reconfiguring, with total configuration times in minutes, rather than hours typically associated with pick-and-place robot. Therefore, Starbugs are excellent candidates for future fibre positioning technology.

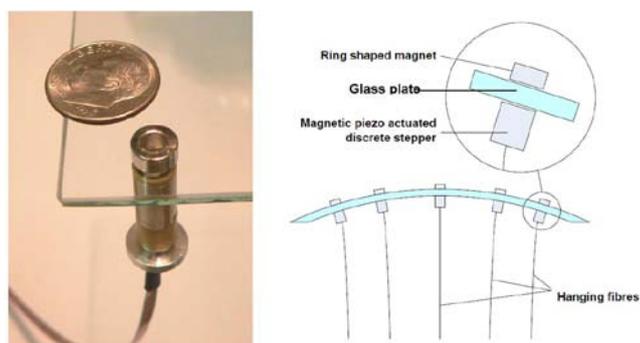

**Figure 1:** *(left) Photograph of the working Starbug prototype; (right) Schematic of the 'hanging Starbug' and 'lift-and-step' concepts. Starlight enters through the ring shaped magnet and is then guided by the fibre to feed the spectrograph. The Starbugs can simultaneously re-position themselves, and move at a rate of a few mm/sec to provide configurations times of minutes.*

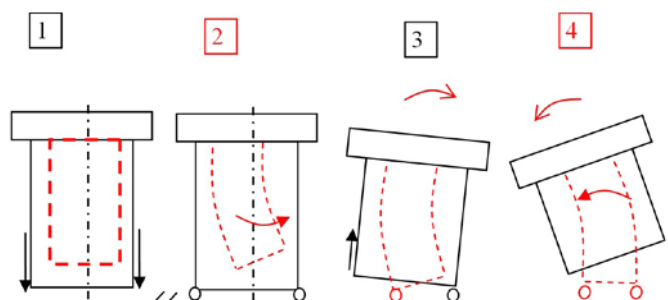

**Figure 2:** *Illustration of the four stages of the 'lift-and-step' concept. Movement of several microns displacements are generated with the application of high voltage waveforms to the inner and outer piezoelectric electric cylinders. Typical voltage waveform amplitude is 120 to 200 V at a frequency of 100 Hz. Each waveform cycle is a single step of approximately 5 microns equating to a typical velocity of 0.5 mm/sec.*

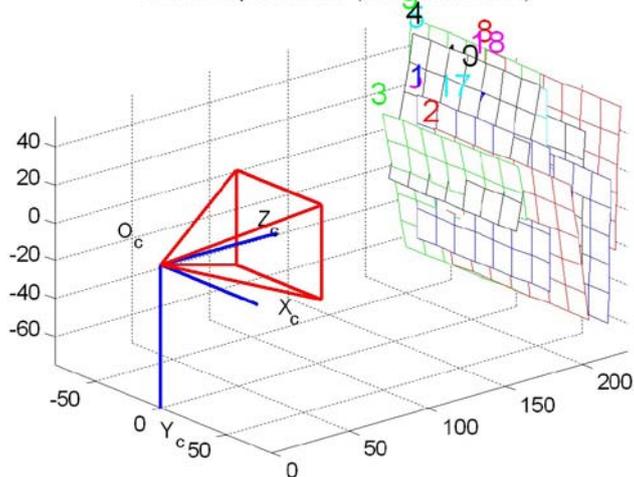

**(a)** *A collection of camera calibration images similar to that shown in (b) are projected in the camera reference frame coordinates (extrinisic). Units are in mm.*

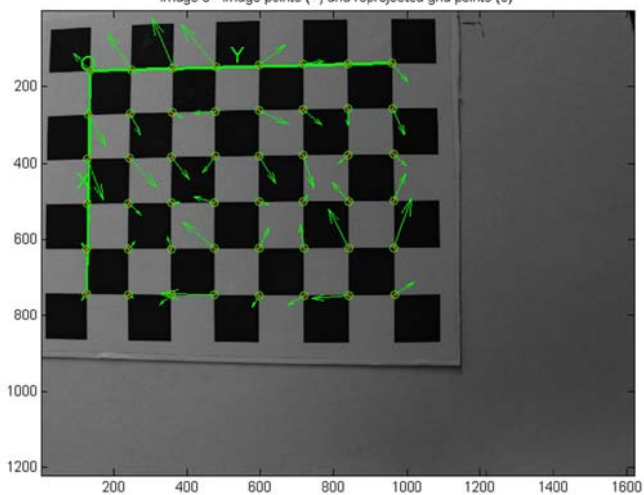

**(b)** *A typical camera calibration image using a chess board pattern. Typically a dozen images are taken at random orientations covering the field of view to determine the camera distortion parameters (intrinsic). Units are in pixels.*

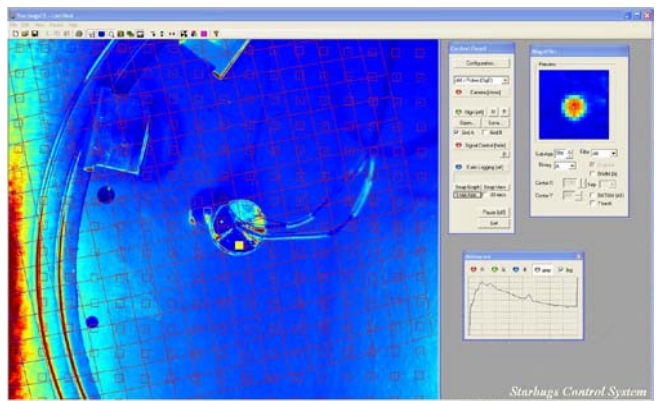

**(c)** *Stabugs Control Software with Starbug*
Figure 3. (a) and (b) Determining the Intrinsic and Extrinsic parameters (camera model) using the Camera Calibration Toolbox for Matlab and a chess board pattern; (c) Starbugs Control Software in development to perform the necessary laboratory metrology.

## Integrated Photonic Spectrograph: First on-sky results

Jon Lawrence, Nick Cvetojevic (AAO/MQ),
Simon Ellis, Joss Bland-Hawthorn (USyd),
Anthony Horton (AAO), Roger Haynes (AAO/AIP)

### The problem

The size of the optical elements in a macroscopic astronomical spectrograph scales with telescope diameter (unless the telescope is operating at the diffraction limit). For large telescopes, this leads to spectrographs of enormous size and implied cost (Allington-Smith 2007). The MUSE instrument planned for the 8-metre VLT (Bacon et al. 2006), for example, will be capable of taking ~90,000 spectra simultaneously over a single field at a moderate spectral resolution (R≈3000). With a volume of ~30 m3, it will occupy the majority of the VLT Nasymth platform. Planned instruments for the 25-metre Giant Magellan Telescope are just as impressive. GMACS, for example, requires a series of very large lenses and diffraction gratings to accommodate a beam diameter greater than 400 mm in each of its eight channels. As with all current astronomical spectrographs the majority of photons collected at the telescope focal plane are 'thrown away' by the instrument. It is clear than alternate solutions to macroscopic element spectrographs are required if future extremely large telescopes are to efficiently collect spectroscopic data from a significant fraction of the telescope field-of-view.

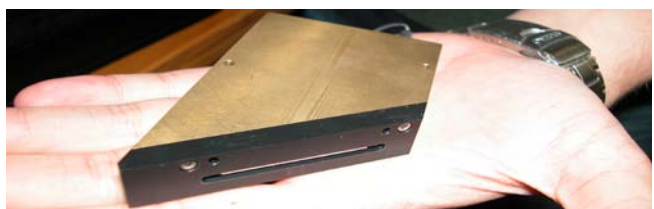

**Figure 1:** *Photograph of the IPS prototype. This device accepts input light from an optical fibre (at the rear) and outputs a spectrum on its front polished face.*

### The solution

The power of astrophotonics to expand the capabilities of astronomical instruments by exploiting photonic technologies has been recently illustrated by a number of demonstration devices (see Bland-Hawthorn & Kern 2009). The Integrate Photonic Spectrograph (IPS) is one such device (Cvetojevic et al. 2009). Three key factors suggest the IPS offers the potential to provide massively multiplexed spectroscopic capabilities for astronomical telescopes:

- *modularisation:* by breaking the system into low-cost modular components, low-maintenance readily- expandable instruments can be contructed with a large number of elements;
- *miniaturisation:* very large multiplexing capacity (e.g. more than a hundred thousand spatial elements) only becomes feasible by significantly reducing the size of each spectrograph element – this is achieved by operating the spectrograph at the diffraction limit with single-mode input fibres;
- *mass production:* by relying on mature lithographic fabrication techniques developed principally by the telecommunication industry we can ensure that instrument costs are not completely overwhelmed by technology development and demonstration costs.

### Arrayed-Waveguide gratings

Several alternative technologies for miniature astronomical spectrographs have been proposed (e.g., Bland-Hawthorn & Horton 2006). Here we have concentrated on the planar arrayed-waveguide grating (AWG). In an AWG spectrometer (Smit & Van Dam 1996), light is coupled from an input fibre, through a free-propagation zone of constant refractive index, into a series of grating waveguides with a constant length increment. The output of these grating waveguides interferes within a second free-propagation zone producing wavelength dispersion. The dispersed spectrum is focussed along a Rowland circle, dependant on the focal length of the output free-propagation zone. Typically, in commercial devices, the output free-propagation zone is interfaced to a series of single-mode waveguide channels that discretely sample the output spectrum.

For astronomical applications, we wish to obtain a continuous spectrum at the device output. For our prototype IPS device, we sliced the chip near the output of the second free-propagation zone, and polished down towards the output focusing surface, so as to remove any waveguide 'stumps' left by the cleaving process. We then packaged the chip to have one single-mode input fibre and an exposed output facet to allow coupling to an external lens or imaging array (see Figure 1).

### Results

After laboratory characterisation (Cvetojevic et al. 2009) we obtained on-sky data with the prototype IPS using the Anglo-Australian Telescope IRIS2 near-infrared imager and spectrograph at Siding Spring observatory during observing runs from 17¬–21 June, 2009. In these experiments, we employed a relay optics assembly to re-image the IPS output focal plane onto the IRIS2 entrance slit, with a plate scale of 1.2 pixels per IPS resolution element. The IPS input fibre was fixed viewing the night sky at a ~45° zenith angle. Due to the lack of strong atmospheric emission lines within the narrowband (i.e., smaller than the IPS free spectral range) IRIS2 filters we used a broadband H filter (1440–1820nm) and a sapphire grism to cross-disperse the IPS output spectrum. In this mode the multiple orders of the IPS which are superimposed on the instrument entrance slit are spread out in a series of bands across the detector array (see Figure 2). This configuration is required as, typical of most commercial devices, the AWG chip has a relatively narrow free-spectral range (~60nm), and operates at relatively high diffraction order (m=27) at its design wavelength (1550 nm).

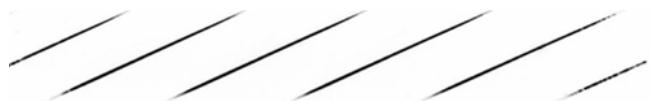

**Figure 2:** Example of a sky spectrum taken during early twilight (i.e., very strong atmospheric emission) with the prototype IPS using the IRIS2 instrument as a cross-disperser. The figure shows ~1000 x 300 pixels. The IRIS2 slit is aligned in the vertical direction. Six spectral orders of the AWG device are being dispersed horizontally.

Using a series of 15min exposures over a period of 12hrs, we were able to detect the night-time atmospheric emission spectrum as shown in Figure 3. In this case, the spectral resolution obtained is a convolution of the IRIS2 resolution with the IPS resolution. To the authors' knowledge, this is the first continuous spectrum to be imaged using an arrayed-waveguide grating spectrometer. The spectrum in Figure 3 shows the detection of upwards of 27 atmospheric OH emission lines across the 380nm filter bandwidth. If we project the dispersed spectrum across the IRIS2 slit axis (i.e., giving the IPS spectral resolution) then the average FWHM across the 27 measured sky lines is consistent with the resolution (i.e., within experimental error) as measured by the laboratory tests (R~2100).

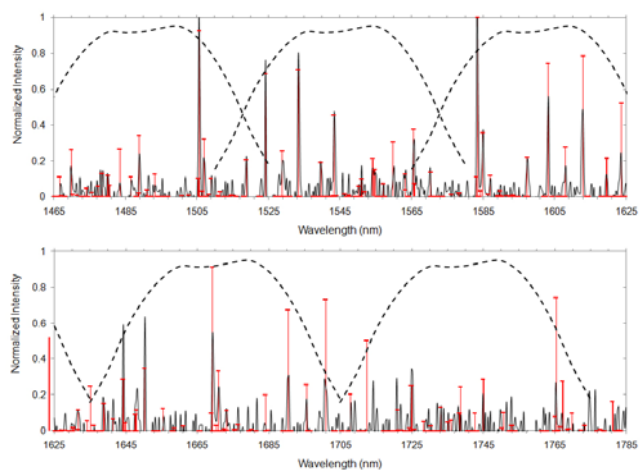

**Figure 3:** The night sky OH spectrum (solid black lines) using the IRIS2 instrument as a cross-disperser for the prototype IPS, superimposed with the theoretical positions and strengths of the OH lines (red), and the diffraction efficiency envelop of the IPS for each order (dashed lines).

### The next steps

While we have demonstrated the feasibility of an IPS for astronomy, there is still a significant gap between this prototype device and a future astronomical instrument based on this technology. There are many questions that must be addressed before it can be determined whether AWG chips are the most appropriate technology for future astronomical spectrographs:

- Can these devices operate efficiently at visible wavelengths?
- How high can we increase the spectral resolving power using this technology?
- What is the most efficient way of sorting out the many multiple diffraction orders inherent in AWG devices?
- What out-of-band blocking filters are required and where should these filters be placed?
- How can very large numbers of AWG chips be stacked?
- How can we efficiently couple multiple single-mode fibres into these devices?
- What is the maximum and optimum number of input fibres per AWG chip?
- How significant is the curved focal plane?
- How can we efficiently couple the output of AWG devices to low-noise cryogenic detector arrays?

The next phase in this project, a collaboration between the AAO, the Astrophysics Institute Potsdam, Macquarie University, and the University of Sydney, aims to address some of these issues. It will involve the construction of a prototype multiplexed IPS device called PIMMS (Photonic Integrated Multimode Micro-Spectrograph). The PIMMS prototype will consist of a series of stacked AWG chips with multiple input fibres and integrated cross-dispersive optics. We aim to obtain an on-telescope demonstration of this device using the AAT in late 2010.

**Send in the clouds: Unveiling the high radio frequency source population with the Gemini poor weather program**

Elizabeth Mahony, Scott Croom & Elaine Sadler (University of Sydney)

While most people hope for clear weather and perfect seeing whilst observing, there is at least one PhD student who is quite happy to take what she can get (especially on an 8m telescope). Over the previous observing semester we have used the Gemini poor weather program to obtain redshifts of radio-loud AGN selected from the Australia Telescope 20 GHz (AT20G) Survey (Murphy et al. 2009).

The AT20G survey is a blind survey of the entire southern sky at 20 GHz down to a flux limit of 40 mJy. This makes it by far the largest and most complete sample of high-frequency radio sources yet obtained, offering new insights into the nature of the high-frequency active galaxy population. Traditionally, radio-loud AGN samples selected below 5 GHz are dominated by objects with powerful jets and lobes stretching well beyond the host galaxy, overpowering the radio emission from the nucleus. Selecting sources at 20 GHz provides a unique sample dominated by objects that derive most of their radio emission from the core of the AGN. Thus completely characterising such a sample will provide insights into the physics of the central engine, providing information about the early stages of active galaxy evolution, which still remains largely unknown (Sadler et al. 2006).

While the AT20G survey is complete at radio frequencies, these data alone are insufficient to constrain models of radio source properties and the evolution of the high radio frequency population. Only ~25% of AT20G sources have redshift information, one third of which are from the 6dFGS survey and the remainder found in NED (the NASA Extragalactic Database). This means that we are biased towards the nearby, optically brighter sources so that our current redshift sample is by no means representative of the full AT20G sample. We are using the Gemini poor weather program to obtain more redshifts with the ultimate goal of having a representative redshift sample to measure the luminosity function of the 20 GHz population and to fully characterise the high radio frequency sky. Selecting optically bright (B<20) sources spread across all RAs makes this sample ideal for the poor weather queue.

This program was allocated 40 hrs, of which over 26 hrs have been observed. We have obtained spectra for 40 objects and obtained reliable redshifts for 35 (88%) of these. The redshifts range from z=0 (due to the presence of a couple of foreground stars) to z=2.9 with a mean redshift of z=0.94. The vast majority exhibit broad emission lines, which is unsurprising as the AT20G survey is dominated by QSOs.

Figure 1 shows some example spectra.

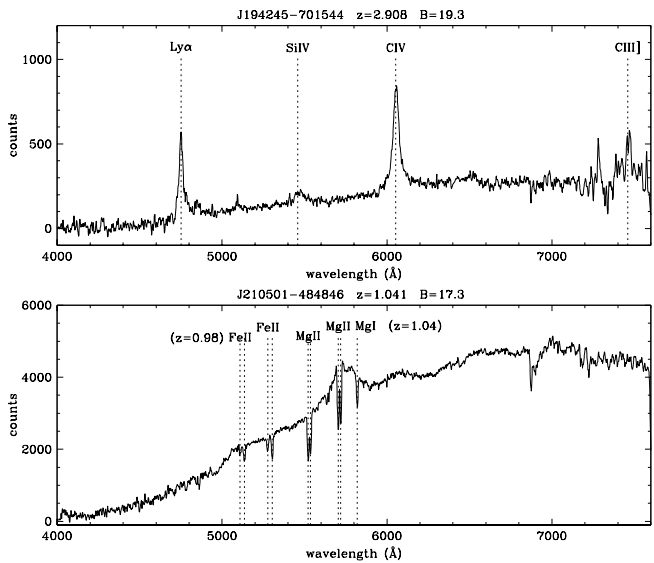

**Figure 1:** *(Top)* A z=2.91 QSO with B=19.3. This source was observed for a total exposure time of 30 mins. *(Bottom)* A z=1.04 QSO with broad MgII emission line and 2 intervening absorption systems along the line of sight, one at the redshift of the host galaxy and one at z=0.98. This source was observed for 12 mins and has a B magnitude of 17.3.

These observations are ongoing, but have already shown that you can still get good data during bad weather. Thanks to Simon O'Toole, Stuart Ryder, Claudia Winge and all the staff at Gemini for helping make this program such a success. If anyone is ever in need of targets to observe when the weather turns bad, I still have plenty more!

**References:**
Murphy T. et al., 2009, MNRAS, in press. (arXiv:0911.0002)
Sadler E. M. et al., 2006, MNRAS 371, 898

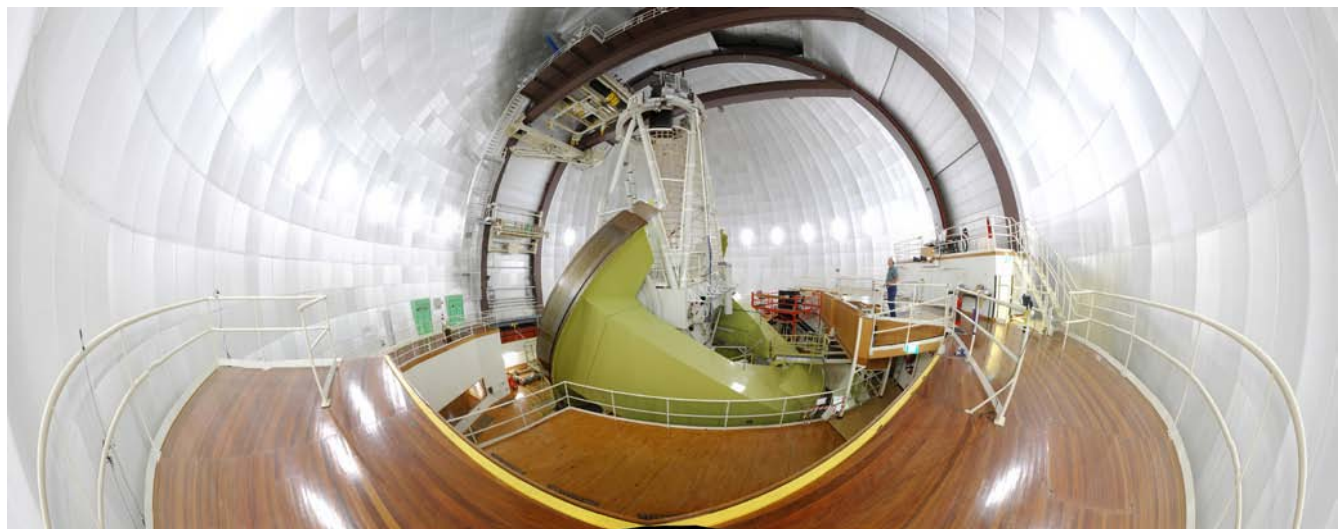

**Anglo-Australian Telescope Dome - Photograph courtesy of Fred Kamphues**

# The MANIFEST Design Study Proposal

Will Saunders, Ian Saunders, Matthew Colless, Jeroen Heijmans, Jurek Brzeski, Andrew Hopkins, Mike Goodwin, Tony Farrell

### Introduction
In November 2008, the GMT board invited design study proposals for first-generation GMT instrumentation. The AAO proposed MANIFEST (The Many-Instrument Fibre System), a general-purpose fibre-positioning system, to feed GMACS (the huge proposed optical imaging spectrograph), NIRMOS (the huge proposed near-infrared imaging spectrograph) and either QSPEC or G-Clef (high resolution optical spectrographs).

MANIFEST allows the spectrographs to observe more objects (because the spectra can be efficiently packed together), over a wider field (the full GMT 20' FoV). Together, these at least double the speed of GMT for survey spectroscopy, making it the fastest of the proposed ELT's. MANIFEST also allows for multiplexed image-slicing, and hence a deployable IFU capability, and/or greatly increased spectral resolution. This makes the GMT uniquely capable of carrying out, e.g., a large Lyman-α tomography survey to determine the structure of the IGM at redshifts 2-3. Fibres also allow OH-suppression, and this will be included into MANIFEST if cost permits, increasing the J and H band sensitivities by 1-2 orders of magnitude.

### Design overview
MANIFEST presents intriguing technical challenges. The focal plane is large, strongly curved, and gravity-varying. Filling the spectrographs requires very large numbers of fibres, which must mimic each spectrograph's standalone input. The available space is limited, and there are severe budget constraints.

**Table 1:** *Deployable unit numbers. GMACS capacities assume nod & shuffle*

|  | Diam. | # Fibres | Spectrograph DFU capacity | | | MANIFEST DFU's | |
|---|---|---|---|---|---|---|---|
|  |  |  | GMACS | G-CLEF | NIRMOS | Starbugs | Pick & Place |
| Single aperture | ~0.75" | 1 | 1200 | 50 | 450 | 1200 Vis/NIR | 50 Vis |
| Image-sliced | ~0.75" | 7 | 500 | 20 | 150 | 500 Vis + 150 NIR | 500 Vis/NIR |
| IFU 2" | ~2" | 37 | 100 | 4 | 30 | 100 Vis + 30 NIR | 100 Vis/NIR |
| IFU 4" | ~4" | 127 | 25 | 1 | 9 | 25 Vis + 9 NIR | 25 Vis/NIR |
| IFU 10" | ~10" | 900 | 4 | – | 1 | 4 Vis + 1 NIR | 4 Vis/NIR |
| Totals |  |  |  |  |  | 2019 | 679 |

### WFC/ADC
It is proposed that MANIFEST share the WFC/ADC system with GMACS. The current WFC/ADC design was intended only for visible use, and has significant throughput losses in the NIR. We have proposed a revised WFC/ADC design, which uses fused silica instead of BK7 for L1, and also has a fused silica 'Dummy ADC' (DADC), which can be substituted for the ADC for NIR (and also UV) use. The all-silica WFC/DADC, and aluminum GMT mirror coatings, would also allow superb UV throughput. This would allow GMACS to be used deep into the UV, either on its own (observing at the parallactic angle) or with MANIFEST (using the deployable IFU's).

Our preferred design is a radical new 'hanging starbug' design, as described in the accompanying article. Starbugs are small autonomous micro-robots, which simultaneously position themselves on the field plate. They allow great flexibility, short setup times, and also micro-tracking (e.g. to correct for changing differential refraction). We propose that the starbugs should hang off a glass dome, removing the need for retractors entirely. The basic layout is shown in Figure 1.

Because the starbug design incorporates untested new technology, we propose to also develop a more conservative 'pick-and-place' option in parallel. This would be based on a commercial H-frame robot, modified only as necessary to accommodate the varying gravity vector and curved focal plane. It would have two interchangeable field plates, each with associated retractors and fibres, like 2dF or OzPoz. The proposed layout is shown in Figure 2.

Buttons, or 'Deployable Fibre Units' (DFUs) would come in a variety of aperture geometries – e.g. single-aperture, image-slicing (to increase spectral resolution), or deployable IFU's of various sizes. The DFU design would make extensive use of hexabundles (ref?) and fibre tapers, which allow simple and efficient designs.

Outputs from DFUs of one type would be assembled into 'modules', terminating in sets of commercial fibre connectors which can be plugged into each instrument as needed.

The starbug design permits huge numbers of DFUs, allowing MANIFEST to fill all of the spectrographs in all of the obviously useful ways. However, a pick & place system is limited to about 600 buttons both by reconfiguration times and by retractor volumes, requiring some DFU rationing. The approximate DFU numbers we are envisaging, and the capacities of the various spectrographs, are as follows:

### Fibre-fed spectrographs
For seeing-limited apertures, any spectrograph fed by GMT is massively oversampled at the detector. This means huge numbers of pixels are needed to achieve adequate resolution and spectral coverage. Image-slicing capability avoids this problem: for example, MANIFEST would be able to feed dedicated, gravity-invariant, fibre-fed 3-armed spectrographs, with 4Kx4K detectors, with 2Å resolution and excellent efficiency. Although many such spectrographs would be needed, the overall cost could be much less than a large imaging spectrograph.

### Performance
***Spectral resolution:*** Image-slicing improves spectral resolution by a factor ~3 over a seeing-limited slit, for all proposed spectrographs. The maximum resolution is ~1Å with GMACS, ~2Å with NIRMOS, and R~100,000 with G-Clef or QSPEC.
***Throughput:*** The proposed design minimizes fibre length and fibre fore- and post-optics. The overall estimated efficiency for single-aperture modes is shown in Figure 3. These estimates do not include the gain in efficiency for NIRMOS and GMACS that comes from using VPH gratings at their superblaze angle. This gain is 10-20%, depending on wavelength and field position; this means MANIFEST is effectively 'throughput neutral' for most survey work. For image-slicing modes, there is an additional efficiency loss of 10-15%, from the filling factor of the hexabundles.
***Wavelength range:*** The goal is for instrumental emission to remain below inter-OH line sky to the end of the H band at 1.81um. At the blue end, the all-silica optics and very short fibre lengths for GMACS should give excellent throughput, so we intend to design down to the atmospheric limit at 320nm.

### Next steps
The GMT board will shortly decide which proposals will be continued to a full 15-month long concept design study. Hopefully, MANIFEST will be included, to give GMT a unique niche amongst the ELT's, in an area where Australia has unparalleled expertise.

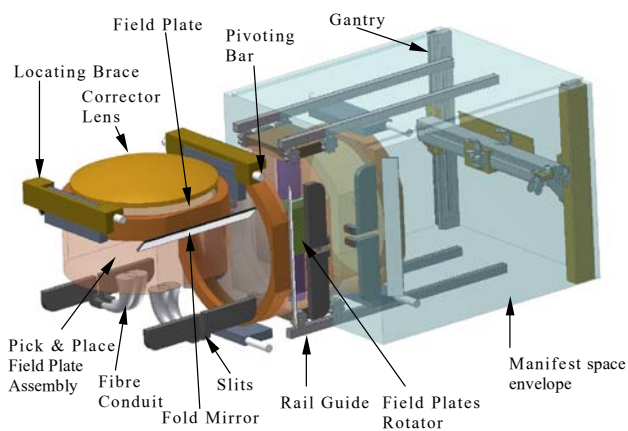

**Figure 1:** *Layout for pick-and-place system. There are two field plates; one is facing the H-frame robot on the right, while the other is shown both in use, and in its retracted position. The available space envelope is also shown. When not in use, everything except the top rails slide back into the envelope.*

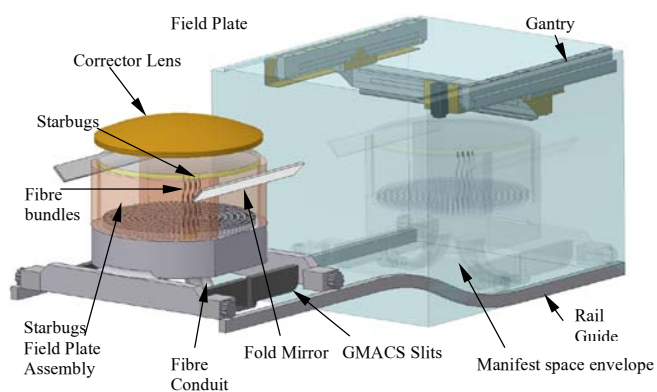

**Figure 2:** *Layout for starbug positioner concept, showing both deployed and retracted positions, and the available space envelope*

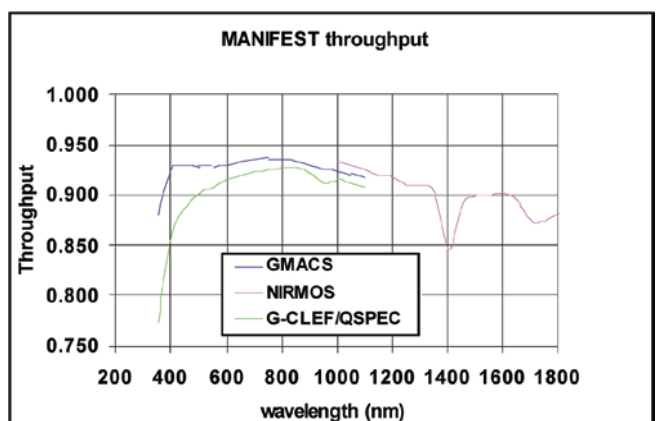

**Figure 3:** *Estimated throughput for single aperture DFUs.*

## Summer students
### Sarah Brough

The AAO runs a twice-yearly fellowship programme to enable undergraduate students to gain 10-12 weeks first-hand experience of astronomical-related research. The current crop of students are from the Southern hemisphere:

Anna Perjma is just about to graduate in a Bachelor of Science Optics and Photonics degree from Adelaide University. Under the supervision of Dr. Jon Lawrence she is characterising the 1x7 and 1x61 photonic lanterns in the infrared. A photonic lantern couples light from a multi mode fibre, and tapers down to individual single mode fibres which then can be analysed using photonic devices such as fibre Bragg gratings. The aims of investigating photonic lanterns are a) to characterise transmission effeciency from the single-mode to multi-mode end and vice versa as a function of focal ratio and b) to observe the near field and far field patterns on the multi-mode end when light is coupled in turn to each of the 7 and 61 single-mode fibres. She hopes to see this exciting new technology installed in the future into telescopes around the world.

Jielai Zhang has just finished the 3rd year of an undergraduate physics/mathematics/engineering degree at the University of Sydney. She is studying the origin of magnetic fields in white dwarf (WD) stars under the supervision of Dr. Paul Dobbie.

Jielai will carry out an analysis that takes advantage of the ~10,000 WDs and predicted hundreds of magnetic white dwarf (MWD) stars targeted for spectroscopy in the Sloan Digital Sky Survey (SDSS), Data Release 7). The SDSS spectra will be used to visually identify WDs and MWDs from a colour selected sample of objects in the survey. The proper motion of these stars will then be statistically analysed to determine whether there is evidence for MWDs to have formed from more massive, short living magnetic stars (Ap/Bp stars are candidates) and thus have smaller scatter in their proper motions.

Ashley Rowlands is half-way through his Astrophysics honours year at Monash University and has one more semester left to complete his Mechanical Engineering degree, at the end he will have a BEng(Hons) and BSci(Hons).

Ashley is working with Professor Quentin Parker on cateregorising objects from the Miscellaneous Emission Nebula(MEN) catalogue; a collection of rejected MASH catalog objects not considered to be true PN.

Jamie Gilbert graduated from the University of Bath, UK, with an MEng in Electronic Engineering with Space Science and Technology in June 2009. He decided to take a year out to travel and assess employment options, but heard about the AAO's Student Fellowship scheme from a former classmate, Talini Jayawardena, who took part last year.

Jamie is working with Michael Goodwin on developing a control system for the Starbugs concept, in which miniature piezo-electric robots position optical fibres by 'walking' them across a telescope's focal plate. So far he has designed and implemented some switching hardware to manage the steering of a Starbug, so from here he will be looking at ways to accurately control the device using software algorithms running on a PC. Ultimately he would like to see several Starbugs moving to a degree of accuracy sufficient for real life use in a telescope.

### What happens when the book talks back?
The Living Library, an unusual outreach experience
Andrew Hopkins

I became a "Living Book" for a few hours on the morning of 17 Nov 2009. The Living Library is an innovative community based initiative which brings people together in a one-on-one conversation. It aims to encourage understanding, challenge negative stereotypes and reduce prejudice. The idea is to provide simple yet powerful strategy for building social cohesion between diverse community members who wouldn't ordinarily meet.

The event was coordinated by Julie Just of Ryde Library, and took place at the Ryde Library. In the two hour session, I was "borrowed" by two library patrons, who each had the opportunity to talk to me about my nominated topic, "Understanding the Universe". The sessions were simply conversations, the kind you would find yourself having about your work with an interested friend or relative. It was striking how different I found this event compared to my other experiences in talking to the public about astronomy. I only spoke with two members of the public, rather than the large audiences more typical of public lectures or school presentations, although I only spent about the same amount of time with each as I would have when giving a presentation to a large group. There was much more interaction, of course, as you would expect. I found this to be the most valuable aspect of the experience, being able to address specifically what the "borrower" wanted to know, and adapting the level of the discussion to the listener's requirements, rather than having an audience passively accepting a prepared presentation.

Despite these differences, a lot of the questions were the same as in many other outreach events, the ones the general public is always intrigued by. "Do you think there's life out there somewhere?" is always high on the list. The sooner we find it and can tell them, "Yes, it's orbiting that little faint star just over there" the easier it will be to deal with this one!

Overall it was an interesting experience. It's probably not the most efficient way to reach a large population among the community, but for the few that do participate there is a solid opportunity to interact very closely and resolve precisely those puzzles that have been bugging them their whole life. It's always educational and rewarding to talk about astronomy to people with different social and educational backgrounds, and to discover what a variety of interests there are. People are smart, and they puzzle over the most unusual things! Being able to help resolve these puzzles is very satisfying.

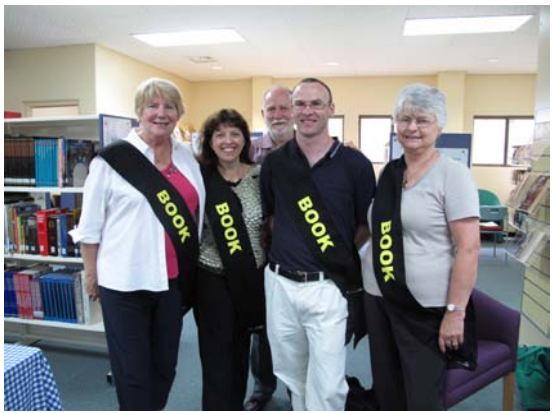

### The 35th Anniversary of the Inauguration of the AAT
Chris McCowage & Helen Sim

I must add a word about the inauguration of scientific equipment in general. Everyone with experience in this matter agrees with the precept that a scientific instrument should be inaugurated at the earliest moment at which those who have provided the money can be persuaded into believing the instrument is finally completed. Otherwise you will have your 'masters' (Hosie's word) on your back all through the difficult period of really completing the thing. It is a concensus among astronomers that you inaugurate a telescope just as soon as possible after you have taken the first aesthetically satisfying pictures with it.

Sir Fred Hoyle, then Chairman of the Anglo-Australian Telescope Board - personal account

Agreement to build the AAT was reached in 1967, with the commencement of the building foundations in October 1970. Ben Gascoigne recounts that early in 1974 it became apparent that the AAT would be essentially completed later in that year. The Board wished to mark the inauguration in an appropriate manner for such an important international scientific project.

Ben Gascoigne took the first photographic plate on the night of 27th April, 1974. This was the cue to inform the Australian Government that planning for the event could commence. Eventually the date was fixed and it was to be a Royal Ceremony. Hoyle was concerned at this prospect, despite assurances that the event would be organised by the Australian Government. The location and the complexity of the building, however soon lead to the decision that the AATB should assume responsibility for inauguration with the majority of the detail planning placed in the capable hands of Board Secretary Doug Cunliffe.

HRH Prince Charles inaugurated the Anglo-Australian Telescope on October 16th, 1974. That day was the culmination of the ambitious project to build the first truly international observatory, which would serve the astronomical communities of Australia and the United Kingdom. In addition to the official party, the event was attended by about 500 invited guests among whom were significant Australian and British scientists, politicians and community members including the then Australian Prime Minister Gough Whitlam and the British High Commissioner Sir Morrice James. The order of arrangements reveals great efforts to observe formality and protocol while accounts by attendees give an insight into the level of preparation required along with the stresses, strains and lighter moments.

The Anglo-Australian Observatory celebrated the thirty fifth anniversary of the inauguration on October 16th, 2009 with a private and a public event.

In the morning, the Siding Spring Community met in the Visitors' Gallery of the AAT to see the unveiling of a plaque commemorating the anniversary. This plaque sits alongside the one unveiled in 1974 by Prince Charles at the telescope's inauguration.

Fred Watson spoke first setting the historical context, inviting the audience to transport themselves to that day thirty five years ago when the inauguration was acknowledging the courage, imagination and commitment shown by so many in so many ways, all inspired by the project and driven by a universal sense of purpose.

AAO Director Matthew Colless recounted the telescope's long list of scientific and engineering achievements and thanked the past and present staff for having made it all possible.

Professor Colless stated that the inauguration in 1974 was "A great day for Australian and British Astronomers" and that even the famous astronomer Sir Fred Hoyle, then chairman of the Anglo-Australian Telescope Board, would have had difficulty imagining the "remarkable things that this telescope has discovered".

Illustrating the role that the AAT has taken in developing the careers of numerous Australian and British astronomers, he revealed that he'd first visited the telescope as a teenager, soon after it was built. His association with the AAO had commenced as a vacation student with the AAO. Then, as an astronomer based in Canberra, he'd used the telescope for many years for his scientific research, finally becoming AAO Director in 2004.

Professor Colless commented on the excellent and productive partnership between the UK and Australia and stated, "We are sorry that they [the British] are going because all of those achievements recounted earlier were the result of the fruitful UK and Australian collaborations, but very happy that the AAO will be continuing and be well supported by the Australian Government".

Councillor Peter Shinton, Mayor of Warrumbungle Shire Council, recalled the construction of the telescope and spoke of how it had become part of the life of Coonabarabran. He then congratulated the AAO and its entire staff on the success of the AAT over the past 35 years.

The unveiling ceremony was followed by a lunch for the staff organised by the Siding Spring Social Committee. This was originally to have been a barbecue but had to be held indoors because of the strong winds. Remarkably the wind had also been extremely strong on the inauguration day in 1974.

In the afternoon, the AAO formally presented an armillary sphere sundial to the Warrumbungle Shire Council on behalf of the community. The gift acknowledges their role in the planning, construction and ongoing support that have contributed to the success of the AAO. This has included helping to protect the night sky by assisting in controlling light pollution around Siding Spring Observatory. The sundial has been permanently mounted adjacent to the Coonabarabran Court House.

Aboriginal elder of the Gamilaraay people, Bill Robinson, gave the welcome to country. Professor Colless spoke on the importance of the relationship between the AAO and the Warrumbungle Shire Council and community. This ranged from involvement of the council in the planning for and construction of the AAT and supporting infrastructure, such as roads and water supply, to the ongoing protection of the famous dark night skies so important to the success of the AAT. He also spoke of the importance of having a welcoming and vibrant community for the local AAO staff and their families.

Councillor Peter Shinton accepted the gift on behalf the community and spoke of the impact of the decision to site the AAT on Siding Spring both at the time and in the long run as an important part of the cultural and economic life of the shire.

Many former AAO staff and community representatives came to this event and to the afternoon tea that followed in the Shire hall.

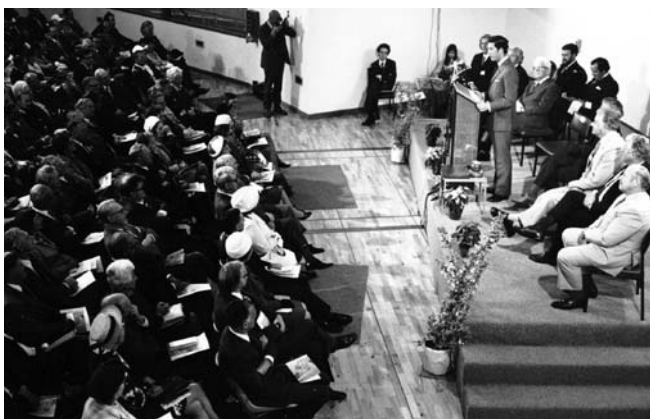

Prince Charles addresses the invited guests on the main floor of the AAT Official Party from left to right:

**Professor Joe Wampler Director**, Anglo-Australian Telescope
**W.L. Morrison** Minister for Science
**Sir Mark Oliphant** Governor of South Australia
**His Royal Highness Prince Charles**
**Sir Fred Hoyle** Chairman of the AATB
**E.G. Whitlam**, Prime Minister
**Sir Morrice James** British High Commissioner
**J. Waddy** Minister representing the Premier of New South Wales

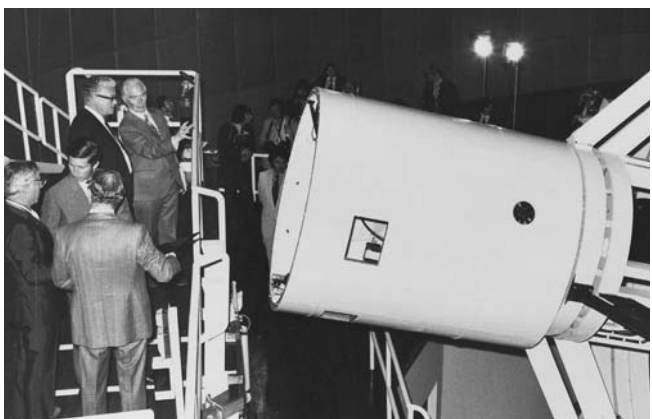

HRH Prince Charles talking with Hoyle and Gascoigne; Robins and the British High Commissioner, Sir Morrice James, are behind them.

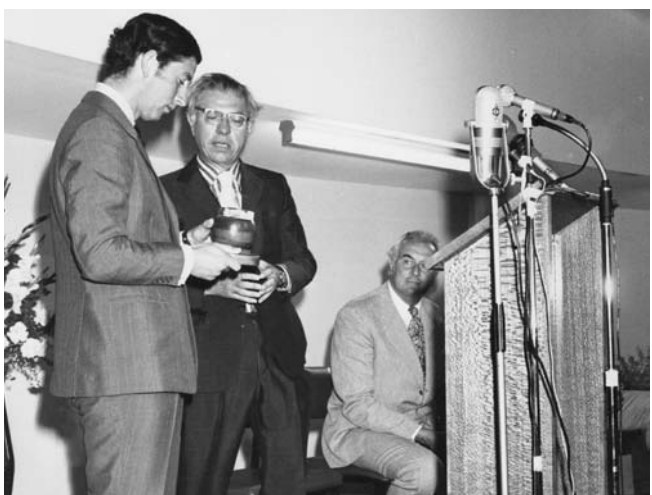

Sir Fred Hoyle presenting HRH Prince Charles with a gift at the inauguration as E. G. Whitlam looks on.

**References:**
A detailed history of the AAO up to 1990 is given in The Creation of the Anglo-Australian Observatory by S.C.B. Gascoigne, K.M. Proust and M.O. Robins, published by Cambridge University Press.
The Early History of the Anglo-Australian 150-inch Telescope (AAT) Bernard Lovell
Q. Jl R. astr. Soc. (1985) 26, 393-455
The Anglo-Australian Telescope by Fred Hoyle, published by University College Cardiff Press

## New AAT proposal form comes into use for Semester 2010B

Heath Jones and Scott Smedley

Over the past few years the AAO has been active in compiling and maintaining a database of AAT use, as part of the Observatory's AAOSTATS system. These data allow the AAO to monitor AAT usage patterns in terms of telescope and instrument demand, oversubscription rates, and the mix of Australian in international participation among many other things. Key AAT stakeholders such as governing bodies and the user community (through its time allocation and user committees) use these statistics to obtain an ongoing picture of how our telescope is used. The information is also used by the AAO itself for strategic planning to ensure the AAT's continuing competitiveness in the era of 8 and 10-m class telescopes.

With the AAT becoming an Australian entity in July, several sections of the old LaTeX AAT proposal form have become redundant, prompting a review of the proposal process. Motivated by this and the long-standing need to better-integrate observer statistics with AAOSTATS, the AAO has moved to a new online AAT proposal form.

Starting with the Semester 2010B round, the AAO will use the new online web form (Figure 1) for all AAT proposals in place of the old LaTeX forms. The LaTeX forms will not be accepted in 2010B nor in subsequent semesters. All applicants for AAT time (both old and new) are strongly urged to familiarise themselves with the new arrangements well before the proposal deadline (15 March 2010). In particular, prospective AAT users should check that both their names and institutions are listed in our database before setting out to apply, and to contact the AAT Technical Secretary (Heath Jones, aatac@aao.gov.au) if they are not. Regular AAT applicants should find little difference in the content of the old and new forms.

Further information about applying for AAT in Semester 2010B can found at http://www.aao.gov.au/astro/applying.html. Specific information about the new AAT proposal form can be found on its help page: http://www.aao.gov.au/astro/apply/ATAC-submit.html.

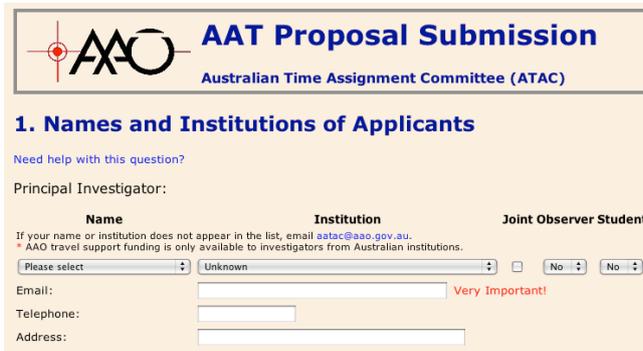

**Figure 1:** *Out with the old and in with the new: a portion of the new AAT proposal form that will come into effect for the Semester 2010B proposal round.*

## Changes to time accounting on the AAT

Matthew Colless and Heath Jones

With the complete UK withdrawal at the conclusion of Semester 10A (31 July 2010), the method for allocating AAT time along needs to be modified. The method used for Semesters 2008B to 2010A inclusive saw proposals draw time from the Australian and UK partner shares in proportion to the nationality mix of the applicants. A detailed description of that method was given in the AAO Newsletter 108, (August 2005).

The new time allocation formula closely follows the spirit of the original method in its use of two parameters to balance rightful AAT share with high-quality science:

1. A specified fraction, $f_O$, of open-access time, taken out of Australia's (otherwise unconstrained) share, $f_A$, such that $f_A + f_O = 1$. The purpose of the open-access share is to foster international collaboration, as well as allowing high-ranking but internationally-dominated proposals to also use the AAT.

2. A super-majority threshold, $M$, used to measure the proportion of Australian involvement in a proposal. The purpose of the super-majority threshold is to ensure that time is awarded largely according to Australia's funding of the AAT's operation while still allowing (and even encouraging) some level of collaboration.

The new time allocation procedure starts, (as it always has), with the Australian Time Allocation Committee (ATAC) ranking all proposals by scientific merit, without regard to the nationality of the applicants. The AAO's scheduler then proceeds to allocate programs in rank order, according to the following prescription:

- Initially the scheduler draws down the Australian and Other time-shares in proportion to the fraction of such proposers on each program. A proposal that is awarded N nights and has A Australians and O Other proposers counts as $N_A$ Australian nights and $N_O$ Other nights, where $N_A = N \cdot A/(A+O)$ and $N_O = N \cdot O/(A+O)$.

- If the Australian share is exhausted first, the remaining proposals are awarded time as ranked, regardless of nationality.

- If the Other share of AAT time is exhausted first, the remaining proposals are only awarded time from the residual Australian share if (i) they have an Australian PI and (ii) they meet the super-majority criterion – i.e. if the fraction of Australians is greater than or equal to the super-majority threshold, $A/(A+O) \geq M$. The exception to this is when there are no qualifying proposals that can make use of the remaining time due to observational constraints.

In Semester 2010B ATAC will adopt a Australian fraction of 70%, an Other fraction of 30% (including the OPTICON transnational access program), and a super-majority threshold of 67%. These fractions reflect AAT demand (based on nationality) over the past few semesters and are supported by the AAO User's Committee and AAT Board. They are not necessarily fixed and will be reviewed in the context of AAT demand over future semesters.

The new method accounts more precisely and straightforwardly for Australian and Other access to AAT time than other procedures that attempt to classify each proposal as either 'Australian' or 'Other'. It also provides a strong bias towards Australian-led and -dominated proposals once the Other share is fully subscribed, without fixing that as a strict cap (or, conversely, strictly guaranteeing the Australian fraction). At the same time, it ensures that smaller programs led by Australians are not compromised if Australian-dominated Large Observing Programs are highly ranked and take up a substantial fraction of the Australian share (currently about 30- 40% of AAT time is allocated to Large Observing Programs). Finally, it provides a strong incentive for Other proposers to collaborate with Australians, particularly if they are seeking a substantial amount of AAT time.

## Large Observing Programs on the AAT
Request for proposals Semester 10B

A new Request for Proposals is being issued for major new observing programs to commence in semester 10B (August 2010 to January 2011) or semester 11A (February 2011 to July 2011). The community's attention is drawn to the AAOmega spectrograph, whose use in the current WiggleZ and GAMA Large Programs is expected to end in Semester 10A, opening up new opportunities for future use of this instrument.

The AAO encourages ambitious large programs and does not set an upper limit on the fraction of time large programs can be awarded. It is expected that large observing programs will be awarded at least 25% of the available time on the AAT; this fraction of time will be reviewed in subsequent semesters.

Proposers are encouraged to form broad collaborations across the Australian and international communities. The status of current large observing programs can be found at http://www.aao.gov.au/AAO/astro/apply/longterm.html.

All proposals will be evaluated by the Australian Time Allocation Committee (ATAC) that replaces the former Anglo-Australian Time Allocation Committee (AATAC) for AAT proposal rankings. Proposals should use the new ATAC AAT proposal form (with non-standard page limits for the Scientific and Technical justification). ATAC will award time based on considerations including the relative scientific merit and impact of the large programs and standard programs, the quality of the management, publication and outreach plans, and the phasing of programs to provide a steady rollover of large programs for the longer term.

**Proposals for large observing programs should be submitted to ATAC by the standard proposal deadline of 15 March 2010.**

**Proposers should consult the Request for Proposals for more information about the process:** http://www.aao.gov.au/AAO/astro/Large_Programs_RfP_10B.pdf

**Anyone considering submitting a large program proposal should contact the AAO Director (director@aao.gov.au) in advance to discuss their plans**

## Separate OPTICON TAC for AAT Semester 2010B

The AAO participates in the OPTICON Transnational Access Program (2009-12) providing travel-related funding to AAT users from EU Member (and Associated) countries. Under the current OPTICON FP7 agreement (2009-2012), AAO telescopes can carry up to *10 OPTICON nights per semester (on average)*, and possibly more, subject to availability of funds.

Starting in Semester 2010B, the OPTICON consortium will operate *a separate Time Allocation Committee*, (distinct from the Australian Time Allocation Committee, ATAC), with an *earlier submission deadline (1 March 2010, 12pm UT)*.

European applicants with OPTICON-eligible programs should be submitted to the OPTICON TAC by March 1 and *not* ATAC.

The AAT OPTICON nights (up to a maximum of 10 nights in 2010B) are top-sliced from the available time but counted as part of the Other (non-Australian) fraction of AAT time (30% overall).

OPTICON-eligible programs that are unsuccessful in securing OPTICON time will automatically be ranked alongside other AAT programs for the remaining pool of AAT time. If they are awarded nights from this non-OPTICON share of nights, they will not be eligible for OPTICON financial support in this round.

Proposers should consult the OPTICON pages at both the AAO and ING for more information:

http://www.aao.gov.au/AAO/astro/opticon.html

http://www.ing.iac.es/opticon/

Information about the 2010B OPTICON proposal call can be found here:

http://www.astro-opticon.org/fp7/tna/opticon_call_2010b.html

Heath Jones (AAT Technical Secretary, aatts@aao.gov.au)

## Comments from the Chair of the AAO Users Committe
Michael Brown

The Anglo-Australian Observatory Users' Committee (AAOUC) provides advice, recommendations and feedback to the AAO director and AAT board on a range of issues, including maximising AAO science, instrument development and telescope operations. These issues are discussed twice a year; at the AAOUC meeting in mid-July and an early-February teleconference. AAO users are encouraged to contact the AAOUC if they wish to provide feedback on any aspect of the AAO, and can do so anonymously if they wish.

During the July 2009 meeting of the AAOUC a number of issues important to AAO users were discussed. This included occupational health and safety issues at the AAT, which are being upgraded to comply with 21st century standards. The development of HERMES and future AAT instruments was also discussed, along with the need to engage the astronomical community in these developments so AAT science is maximised.

Long-standing issues for the AAO and astronomical community were also raised, including the participation of women in astronomy and having the correct balance of members on the TAC.

At the time of writing, the next AAOUC teleconference is less that 2 weeks away, and the AAOUC is looking forward to discussing developments on each of these issues and addressing concerns raised by AAO users.

### AUSGO CORNER
Stuart Ryder (Australian Gemini Office, AAO)

**Semester 2010A**

In this semester we received a total of 32 Gemini proposals, of which 17 were for time on Gemini North, 4 were for exchange time on Keck or Subaru, and 11 were for time on Gemini South. The oversubscription factor for ATAC time on Gemini North (including Keck and Subaru exchange time) was down from last semester's record of 4.5, to a still very healthy 3.6. The demand for Gemini South time was only half that of 2009B, with an oversubscription factor of 1.24. For the first time in a long while proposals were received for both mid-IR instruments, and the first Australian proposal for non-campaign NICI time was also received. At the ITAC meeting Australia was able to schedule 18 programs into Bands 1–3 (including both mid-IR programs and the NICI program), plus one more in the Poor Weather Queue.

For Magellan we received a record 11 proposals, resulting in the highest-ever Magellan oversubscription of 4.0. The availability of three new instruments in 2010A (the optical wide-field mosaic camera Megacam; the near-IR imager and multi-object spectrograph MMIRS; and the Planet Finding Spectrograph PFS) has further stoked interest in using Magellan. Ultimately time was awarded to two proposals using MIKE, and one each for MMIRS and the PFS.

**Gemini School Astronomy Contest**

As its contribution to the International Year of Astronomy in 2009, AusGO ran a contest for Australian high schools to win one hour of time on Gemini South to image their favourite target in multiple filters with GMOS. The winning entry from Daniel Tran, a Year 10 student at PAL College in Cabramatta, NSW, proposed to observe the planetary nebula NGC 6751, nicknamed the "Glowing Eye". The observations for Daniel's program included observations in the narrow-band Hα, [O III], and [S II] filters and were executed in the Gemini South queue on the night of 25 July 2009. After standard processing, the 3 filter images were colour-combined by Dr Travis Rector at the University of Alaska Anchorage. The awesome result shown on the front cover of this Newsletter was first unveiled at a special assembly of PAL College on 23 September 2009 by the contest organiser, RSAA-based Deputy Gemini Scientist Dr Christopher Onken (Figure 1). Dr David Frew, a Macquarie University planetary nebula expert was also in attendance to explain the scientific significance of Daniel's observations. The winning image and story behind the contest have been featured as a Gemini websplash (http://www.gemini.edu/node/11329), as well as in the December 2009 issue of GeminiFocus, and on the cover of the January 2010 issue of Australian Sky & Telescope.

In order to build on the legacy of this IYA event, AusGO intends to run a similar contest in 2010. We gratefully acknowledge the support of ATAC, which has again allocated us one hour of Gemini South time in Semester 2010A to make this possible. Details of this year's contest have been posted on the contest web site (http://ausgo.aao.gov.au/contest/) and the deadline for entries from Australian high school students is Friday 7 May 2010.

**AGUSS**

Each year since 2006 AusGO has offered talented undergraduate students enrolled at an Australian university the opportunity to spend 10 weeks over summer working at the Gemini South observatory in Chile on a research project with Gemini staff. They also assist with queue observations at Gemini South itself, and visit the Las Campanas Observatory where the Australian Magellan Fellows are based.

The Australian Gemini Undergraduate Summer Studentship (AGUSS) program is generously sponsored by AAL. Once again the selection panel was faced with a difficult task in selecting just two lucky recipients from the 21 excellent applicants. The 2009/2010 AGUSS recipients are Daniel Burdett from the University of Adelaide, and Courtney Jones from the University of Tasmania. Courtney is working with Steve Margheim on a project looking at stellar Lithium abundances in open clusters, while Daniel is working with Gelys Trancho on photometry, ages and metallicities of globular clusters. They are due to present the results of their research via video link from Chile to the AAO/ATNF summer student symposium in Australia in mid-February 2010.

**Gemini and Magellan Publications**

AusGO maintains a master list of all publications involving Australian authors making use of Gemini (including Keck and Subaru exchange time) and Magellan allocations from ATAC: http://ausgo.aao.gov.au/pubs.html. This list is compiled from Gemini's and Magellan's own master lists of publications, but AusGO encourages all authors to contact us in advance of impending publications, so that we can offer you assistance with a press release.

There were 26 papers published in refereed journals in the 2008/09 financial year based on Gemini data and involving Australian authors, up from 21 in 2007/08. In the same period there were 6 papers published in refereed journals based on Magellan data and involving Australian authors, up from 2 in 2007/08 (recall that Magellan access has only been offered since January 2007).

A recent analysis of Gemini publication statistics for 2005–2008 by Associate Director of Science Operations Dennis Crabtree yielded a couple of notable points. First, although Australia is a 6.2% partner in Gemini, papers based on programs to which ATAC contributed some or all of the observing time made up 8% of all Gemini publications. Most outstanding of all is that Australian Gemini publications have an impact factor (defined as the number of citations to a paper, relative to the median number of citations to all AJ papers in that same year) of 7.3, the highest of all Gemini partners, and more than twice that of US and UK Gemini papers.

**Acknowledgements in Publications**

While on the subject of publications, we would like to remind Australian authors of papers based on Gemini or Magellan allocations of the following acknowledgement policies:

- Acknowledging Gemini: Authors of papers making use of Gemini data are required to include the standard acknowledgement text (http://www.gemini.edu/sciops/data-and-results/acknowledging-gemini) in the Acknowledgements section of their paper, as well as noting their Program ID in the Observations section, to assist us in tracking scientific output of the Gemini telescopes.
- Acknowledging Magellan: Authors of papers making use of Magellan time awarded through ATAC are required to acknowledge this in their papers with:

*Australian access to the Magellan Telescopes was supported through the Major National Research Facilities program of the Australian Federal Government.*

if the observations were taken before 1 January 2009, or:

*Australian access to the Magellan Telescopes was supported through the National Collaborative Research Infrastructure Strategy of the Australian Federal Government.*

otherwise. Additionally, the Magellan Council requests that all papers based on Magellan data include the words "This paper includes data gathered with the 6.5 meter Magellan Telescopes located at Las Campanas Observatory, Chile" as a footnote to the title, and provide them with an ADS reference for all such papers (http://baade-clay.org/content/magellan-publications).

**Future Gemini Instrumentation Priorities**

With the Gemini Planet Imager to be commissioned in 2011 marking the conclusion of the Gemini "Aspen" instrumentation program conceived in 2003, the Gemini Science Committee has begun the process of defining Gemini's future instrumentation needs. Prior to the GSC meeting in Hilo in mid-October 2009, the Gemini user community in most partner countries was surveyed about the relative priorities of a representative set of workhorse instrument concepts including:

- a cross-dispersed high-resolution near-IR échelle spectrograph with R >= 50,000 and simultaneous coverage across the H & K bands;
- a high-resolution optical échelle spectrograph with R in the range 40,000–100,000;
- a medium-resolution 0.3–2.5 µm spectrograph with R ~ 3000–5000 offering optical through near-IR spectroscopy in one shot;
- a small field-of-view near-IR Integral Field Unit with OH-suppressing fibres (building on the AAO's groundbreaking achievements in this area – see page 15 of the Feb 2009 AAO Newsletter);
- build no new instrument, but instead upgrade the existing instrumentation (GMOS, ALTAIR and NIRI), and introduce Ground Layer Adaptive Optics that would provide adaptive optics correction at wavelengths from 0.6–2.5 µm over a 10 arcminute field of view.

An AusGO survey of the Australian community found that 1/3 of the 50 respondents ranked the optical échelle spectrograph as the highest priority; 1/3 ranked the OH-suppressing IFU highest; and the remainder were split fairly evenly amongst the other three concepts. These findings were conveyed to the GSC meeting by Australia's representative A/Prof. Scott Croom. The GSC has now undertaken a second round of community surveys focusing on capabilities rather than instrument concepts, and the results are now being assessed by the GSC in formulating its recommendations to the Gemini Board's May 2010 meeting. AusGO and the Australian Gemini Steering Committee appreciate the extensive and thoughtful input from Australian Gemini users into this process.

**GMOS-South CCD Upgrades**

In November 2009, Astronomy Australia Ltd (AAL) issued a call for Expressions of Interest (EoIs) from Australian institutions in developing an Investment Plan for astronomy infrastructure, to be funded from a $10M Education Investment Fund grant to AAL. Of the 18 EoIs received, 8 were selected for further consideration and development of a project plan. One of these proposals, led by Prof. Brian Schmidt at RSAA proposes replacing the nearly decade-old E2V CCDs in GMOS-South with modern E2V devices. In the past 2 years, 2/3 of Australia's time on Gemini South has made use of GMOS-South. The current CCDs suffer from poor quantum efficiency and severe fringing in the red, and subtle cosmetic defects that make nod-and-shuffle observations difficult over large fields-of-view. The replacement E2V CCDs would be of identical size to the current ones (making their integration into GMOS-South relatively straightforward) but offer quantum efficiency improvements of at least 10% in the blue, and up to a factor of two in the red.

Gemini have already contracted the Herzberg Institute of Astrophysics in Canada to integrate new CCDs from Hamamatsu with even better quantum efficiency in the red (but a larger physical footprint than the E2V CCDs) into GMOS-North, but at this time they lack the resources to perform a similar upgrade of GMOS-South. In exchange for Australian funding of the GMOS-South CCD upgrade, Gemini would count this as a "down payment" on our contributions to the Gemini instrumentation program, and/or offer Guaranteed Time on the upgraded instrument. AusGO fully endorses this opportunity to further strengthen Australia's engagement with Gemini, and which will pay immediate dividends for Australian astronomers and all GMOS-South users.

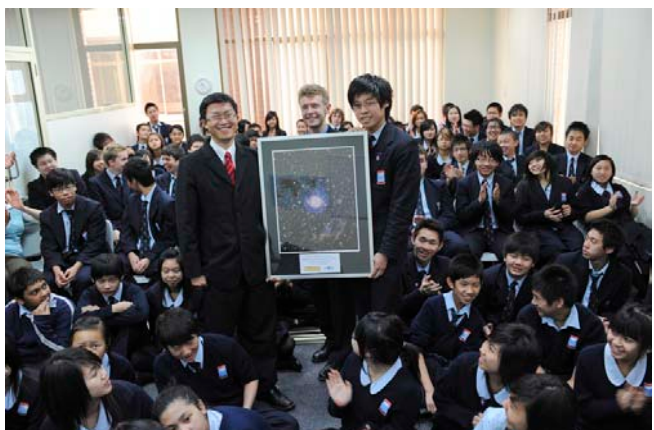

**Figure 1:** *Gemini School Astronomy Contest winner Daniel Tran (right) holds a framed print of his picture, presented to him and his teacher Mr David Lee (left) by Deputy Gemini Scientist Dr Chris Onken (centre) in front of Daniel's classmates from PAL College (photo David Marshall Photography).*

### Letter from Coonabarabran
Rhonda Martin

The 35th anniversary of the inauguration of the AAT by HRH Prince Charles in 1974 dawned fine and clear. It was a busy day for all concerned, with a small ceremony in the AAT Vistors' Gallery when after the mayor of Warrumbungle Shire, Peter Shinton, Astronomer-In-Charge Fred Watson, and the AAO Director, Matthew Colless spoke a few words, a celebratory plaque, suitably garbed in a purple curtain, was revealed. Staff and visitors crowded into the Gallery for this important event.

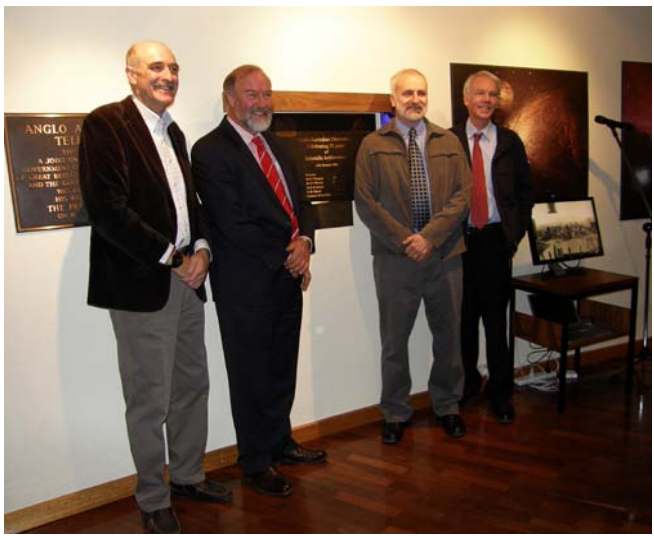

Then it was down to Coonabarabran for all concerned for the official opening of a sundial, donated by the AAO to the people of Warrumbungle Shire in appreciation of their assistance over the years in making the observatory such a success. The sundial itself, sitting on a very natty sandstone plinth, is seated on the footpath outside the courthouse, also made of sandstone. The current Operations Manager, Doug Gray, being a bit shy, it fell to the previous incumbent, Chris McCowage, to explain how the sundial worked and how to adjust for daylight saving. Then we adjourned to the Shire Hall for a very tasty afternoon tea. All in all, a very successful day.

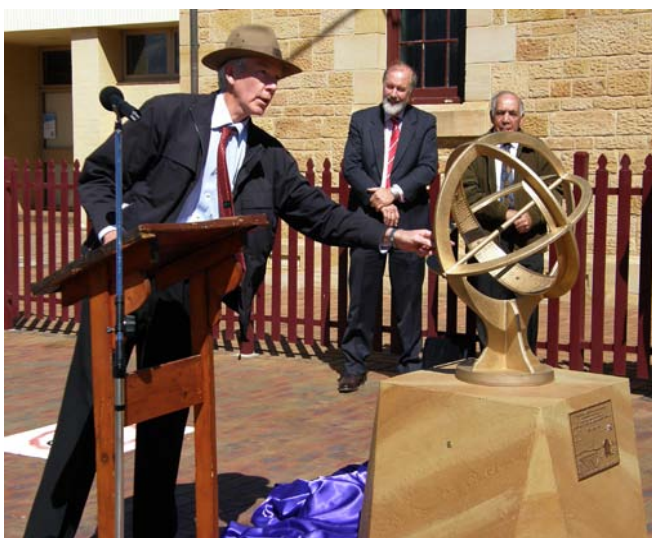

What a wonderful end to the year we had with, finally, rain commencing on Christmas Day and again at New Year. The locals were dancing in the streets – flooded – and although the weather was not condusive to any observing we heard no complaints from observers. It was the best Christmas gift we could have had and because of it we have frogs again sounding their mating calls and cicadas shrilling, the sounds of summer.

Mick Kanonczuk, from the AAT mechanical section has been off for some time after an operation on his back. During his absence his place has been taken by Greg Canham. Welcome Greg and thank you for your assistance.

### EPPING NEWS
Sarah Brough

This Newsletter we have the thrill of being able to report an Australia Day honour for our very own Fred Watson. More details are included in a Newsletter article further on.

We also have had sad news with the tragic loss of our Librarian of 14 years, Sandra Ricketts, to leukaemia. A tribute from her husband, Brian, is also included in the Newsletter.

We are also organizing a large symposium to be held in Coonabarabran in June to "Celebrate the AAO: Past, Present and Future". The Symposium Announcement is included in the Newsletter.

As always since the last Newsletter we have farewelled a number of colleagues and welcomed others.

Roger Haynes, latterly Head of Instrumentation Science, has left after 11 years to head up the Astrophysical Institute Potsdam (AIP) Instrumentation group. Dionne Haynes (formerly James) has joined Roger there after many varied roles at the AAT including UK Schmidt Telescope Observer IT Support Officer, and most recently Instrument Scientist. William Rambold, Instrumentation Group Manager for 3 years has also left for the AIP. Don Kingston has retired after 12 years at the AAO, after 8 years as Accountant and most recently as Building Facilities Manager. Devika Hewawitharana, Accounts Clerk, May 1998 to Nov 2009. Paris Constantine has also moved onto newer climes after assisting in the development of the WFMOS build proposal. Cathy Café left after 1 year as Payroll Officer.

New faces at the AAO include Allan Lankshear, a Mechanical Engineer, a new Accounts and Admin Officer, Vanessa Bugueno and a new Librarian, Phillipa Morley. We also have the first Magellan Fellows fresh from their 2.5 years in Chile, now working in Australia. Ricardo Covarrubias is based at the AAO whilst David Floyd is based at the University of Melbourne. Professor Guy Monnet, the director of the instrumentation program at ESO for many years is visiting the AAO for a year

Ricardo Covarrubias: In December of 2009, I moved to Australia to continue the last 15 months of my Magellan Fellowship at the AAO. I grew up in La Serena, Chile, where I used to see Cerro Tololo Observatory when we had family trips to the Elqui Valley. This made me wonder about the stars and the universe and I slowly became interested in Astronomy. I completed my undergraduate studies in Physics at the Pontificia Universidad Catolica de Chile and then I went back to La Serena to work as a Research Assistant at Cerro Tololo Inter-American Observatory. In 2000 I moved to Seattle and completed my PhD at the University of Washington. I was offered the Magellan Fellowship and moved back to La Serena in May 2007, where I spent 2.5 years working at Las Campanas Observatory.

My scientific interests focus mainly on Core Collapse Supernovae. I am interested in the role the abundance of the supernova site plays in the observable properties of Supernovae Type II, and I am in the process of expanding this research to Supernovae of Type Ib and Ic. I am also collaborating with the ACS Nearby Galaxy Survey (ANGST) group where we are undertaking a project to increase the number of core-collapse SNe with precursor mass estimates, using well-established techniques of stellar population modeling to age-date the stars surrounding the site of the SN. I am looking forward to a great experience during my stay at AAO.

David Floyd: I am a Magellan Fellow hosted by the University of Melbourne since September 2009. I spent the previous two and a half years supporting observations at the Magellan Telescopes in northern Chile. I obtained my PhD from the University of Edinburgh in 2005, and worked at the Space Telescope Science Institute on my thesis and related projects from 2004 to 2007. My research focuses predominantly on quasars, from their accretion disks out to their host galaxies and large-scale environments. I like cheese. And wine. Not necessarily in that order.

Professor Guy Monnet, the director of the instrumentation program at ESO for many years is visiting the AAO for a year: Over some 48 years as an astrophysicist, my main themes have been first large-scale galactic gas dynamics, then dynamics of their stellar components. I had my full share of managerial responsibilities, starting in 1971 as Director of Marseilles Observatory, then successively as Director of Lyon Observatory, Executive Director of CFHT and finally Division Head at ESO. I've always kept a strong involvement in cutting-edge instrumentation via inter alia the CIGALE scanning Fabry-Perot spectrometer and the TIGER integral field spectrometer. As head of ESO Instrument Division I've supervised all VLT & La Silla instrumentation: that included bringing AAO on board for the building of the Oz-Poz robot for the very successful VLT FLAMES facility. I'm very much looking working now at AAO on the exciting HERMES and MANIFEST developments.

## The Anglo-Australian Observatory Symposium 2010

"Celebrating the AAO: Past, Present, and Future"

In 2010 the Anglo-Australian Observatory will see both its 36th anniversary and its transition from a joint UK and Australian facility to a wholly Australian entity, the Australian Astronomical Observatory. This transition will occur on 1 July 2010.

To celebrate the past successes and the exciting future of the AAO, we are holding a Symposium with the theme "Celebrating the AAO: Past, Present and Future". The meeting will be held in Coonabarabran over 21-25 June 2010, with around 100 attendees expected.

Details of the Symposium can be found on the conference web page at: http://www.aao.gov.au/AAO2010/

Details of registration and payment are now available from the web page. Registration is now open, and the deadline for registration is Thursday 1 April. See: http://www.aao.gov.au/AAO2010/register.html

Details of transport, accommodation, and other events are also given on the registration page. Please indicate these when you register. Buses will be available for travel from Sydney to Coonabarabran and back, stopping for lunch, and optional wine-tasting, along the way. When you register, please indicate whether you require a seat on a bus.

Please make sure to pay for the various travel options and activities, in addition to your registration fee, when you visit the payment page at:
http://www.aao.gov.au/AAO2010/payment.html

Conference delegates are reminded that accommodation in Coonabarabran is limited, and you should book accommodation soon. We have arranged for block bookings in a number of local venues, detailed on the web page. You should indicate that you are attending this event when you book.

The program will be split into three broad sections. 1: Historical aspects of, and events surrounding, the AAO. 2: Current developments at the AAO and with the AAT, including exciting new scientific results and the imminent transition. 3: Future plans for the AAO.

We hope you will be able to join us for what we expect to be a fascinating and enjoyable celebration covering historical overviews of the AAO, ongoing research, and eagerly anticipated developments.

**Keynote Speakers:**

| | |
|---|---|
| Ian Corbett | Warrick Couch |
| Roger Davies | Michael Drinkwater |
| Simon Driver | Richard Ellis |
| Bob Frater | Ken Freeman |
| Peter Gillingham | Karl Glazebrook |
| Malcolm Longair | David Malin |
| Donald Morton | Michael Rowan-Robinson |
| Elaine Sadler | Jason Spyromilio |
| Chris Tinney | Fred Watson |
| Hermann Wehner | |

**SOC/LOC:**

| | |
|---|---|
| Matthew Colless | Sarah Brough |
| Russell Cannon | Andrew Hopkins |
| Steve Lee | David Malin |
| Stuart Ryder | Chris Springob |

### Obituary for Sandra Ricketts
Brian Ricketts

Some sad news reached the Observatory just as we started bringing this edition of the Newsletter together. The AAO's Librarian and Newsletter Editor of 14 years, Sandra Ricketts, lost her struggle with leukaemia on the 10th January 2010. In order to offer tribute to Sandra's contribution to the AAO we include the words of her husband, Brian, given at her funeral:

I would like to run through the main features of Sandra's life to try to give you some idea of how she came to be the person that we know and loved.

Sandra Denise Clark was born in August 1944 in Belfast. Sandra's father was an engineer from Glasgow who had come to Northern Ireland to work with the linen industry and Sandra's mother was a local girl from Ballymena near Belfast. Sandra was their only child and she grew up in a very small family with no aunties or uncles and no first cousins. Life in post-war Belfast was quite comfortable for the Clarks although shortages and rationing made sure that no child was overwhelmed with material possessions. Roller skating with the other kids in the street and plenty of time to read during a rather strict observance of the Sabbath seemed to be the main things that Sandra remembered about her early childhood.

As the 1950s went on Sandra's parents could see that the linen industry in Ireland was contracting so they started to think about emigration. Initial thoughts were directed at Montreal. Sandra's great aunt who lived in Chatswood, NSW, Australia heard of this and immediately her family organised to sponsor the Clarks to come to Sydney in 1958 and after a month or two they were able to buy a house in the same road.

Fourteen was not an easy age for Sandra to leave the friends and things she was familiar with; for a start her cat Pitch had to be sent to a farm outside Belfast and that really hurt. She started school at North Sydney Girls School which in those days was very Anglo-centric with all the girls speaking proper Australian. Sandra set to work to get rid of her accent but fortunately she was not wholly successful with this and she was left with her unique mixture of Scottish/Irish and Australian accent which has always intrigued people and was always music to my ears.

After school Sandra did a science degree at the University of NSW specialising in maths and statistics. For one of the vacation training periods over the Christmas of 1964 she made the mistake of ticking the box for working in the Electricity Commission headquarters in Sydney. This place was, in fact, a den of predatory male engineering cadets who had just finished their course and who were looking to start their careers as engineers and men of the world. Sandra stepped into a lift in front of one of those males and things moved on from there.

After graduation Sandra worked for a time in the Reserve Bank but following our marriage in May 1967 this was not an attractive option because at that time married women were a class apart in Commonwealth employment. She did a part time Dip Ed at the University of New England, Australia and started teaching maths at various girls' schools. Three years of the early 70s were spent in Liverpool, England then Tony was born back here in 1975 and Vanessa in 1977.

After several years teaching part-time Sandra decided she really would not mind being a librarian. She did a part-time Diploma in librarianship and got a job in the technical department of the National Roads and Motorists' Association. Shelves of car workshop manuals were not exactly what she had in mind but she made the best of it, conditions were good and she got on well with the other staff. However some years later she was thrilled to be appointed as the librarian of the Anglo-Australian Observatory at Epping. The 14 years she was there were easily the happiest and most satisfying years of her working life. The more academic and unworldly atmosphere of the observatory suited her just fine and she had a lot of satisfaction in being one of the very few specialist astronomy librarians in Australia. With her scientific background she felt part of the observatory team rather than being someone who just ran the library. She contributed to two of the international conferences for astronomy librarians that were held in Prague and Cambridge in the US and she co-edited the proceedings of the Cambridge conference. Her logical and orderly mind meant that she had a wonderful ability to find things; things that had been mislaid by the absent minded and the less well organized. This ability was developed over many years on the domestic front but proved to be very useful when dealing with her astronomers.

She enjoyed her regular drives up to the telescope at Coonabarabran to maintain the small library up there. She was a bit of a lead-foot in a car so I was always very pleased to see her home safely. Usually she drove alone but it was a feather in her cap that on one occasion she shared the trip with the observatory's very own convicted felon although it must be said that that particular felon was basically a nice guy and was no real threat to anyone.

Morning teas were a must and the Wednesday lunches were not to be missed either. Sandra made many friends at the observatory and I do not want to start listing people but she especially appreciated her friendship with the two Helens.

Sandra's greatest achievements in life were to bring Tony and Vanessa into the world and support them to become the people we have here now. She knew how to encourage them without trying to live their lives for them. She was thrilled to get Vanessa married off in 2002 and Tony in February last year and warmly welcomed Stephen and Michelle into the family. Needless to say her granddaughter Madeleine has been a source of great joy to Sandra through the whole of the last 9 months.

Left to her own devices it probably would never have occurred to Sandra that sailing would be a good way of spending the time. Perhaps just to please me she agreed to come sailing back in the 60s in a small boat on Middle Harbour. It was a rather unstable boat and we spent a fair bit of time in Middle Harbour. Much later in 2003 we decided to try a larger boat in Pittwater. Sandra came to really enjoy sailing and we became quite a team with Sandra doing most of the helming while I did the ropes.

Depending on conditions, feelings ranged from blind panic in heavy swell near Lion Island to the most blissful peace when overnighting in one of the bays. We both had a lot to learn and on every trip there was sure to be something to take us right out of our comfort zone.

I suppose you could say that sailing was my idea but Sandra's really personal interest was embroidery. It started with classes in creative embroidery more than 20 years ago and she has moved through stages such as antique samplers, goldwork, stumpwork , ribbon embroidery and quite elaborate 3 dimensional things. Sandra has always found embroidery a very peaceful activity and the enjoyment was always in the doing rather than displaying her handiwork. She has been to many classes where everyone is doing the same thing but she had most enjoyment from just sewing in the company of others where each person was just doing their own thing. As you can imagine we have a quite a number of embroideries on our walls so any embroiderers here might like to cast an eye over them after the service. It goes without saying that with a wife so heavily into needlework it has always been a real battle to get a shirt button sewed back on. Shirt buttons are beneath the notice of some needlewomen, it seems.

Sandra has been, of course, the love of my life. It is hard to properly put my strong feelings for Sandra into words and I should not try that here and now. I think that there were two things that were most noticeable to everyone about Sandra. One was that she was simply a very happy and optimistic person and the other was that she had a wonderful welcoming smile for everyone. These qualities and the smile will live on with Tony and Vanessa and I am sure that, particularly when she gets a full set of teeth, Madeleine will carry Sandra's smile even further into the future.

## 'Out of this World' Honour for Australian Astronomer

Fred Watson

Professor Fred Watson, Astronomer-in-Charge of the Anglo-Australian Observatory, and one of Australia's best-known science communicators, has been honoured for his services to astronomy. On Australia Day, January 26, Fred was appointed a Member in the General Division of the Order of Australia.

'It's truly an out-of-this-world experience to find yourself in the Australia Day honours list,' said Fred. 'We live in an era when astronomy and space science are exploding with new discoveries, so it's quite easy to spread the excitement around. This honour reflects the generous support I've had over the years from friends and colleagues in Australia and worldwide.'

Fred has been Astronomer-in-Charge at the AAO since 1995, having previously worked at the Royal Greenwich Observatory and the Royal Observatory, Edinburgh. Acknowledged in professional circles as one of the pioneers of fibre optics in astronomy, Fred is currently Project Manager for the international RAVE survey of a million stars. He holds adjunct professorships in the University of Southern Queensland, Queensland University of Technology and James Cook University.

It is for Fred's popular science that he is best known. His frequent appearances on ABC radio and TV, together with his books, public lectures and astronomy tourism expeditions, have resulted in several awards. They include the David Allen Prize for Communicating Astronomy to the Public, the Australian Government Eureka Prize for Promoting Understanding of Science and the Queensland Premier's Literary Award for Science Writing for his book Why is Uranus Upside Down? In 2004, asteroid 5691 was named 'Fredwatson' in his honour (though he is always at pains to point out that if it hits the Earth, it won't be his fault).

Fred's enthusiasm for linking science and the arts also led to a solo CD, An Alien Like You, featuring some of his quirky science songs. At the other end of the musical spectrum, Fred was librettist for Star Chant, the choral Fourth Symphony of Australian composer Ross Edwards. Following its release on an ABC CD, Star Chant won the APRA Award for Best Choral or Vocal Work of 2008.

Fred was born in Yorkshire, but is proud of his Australian citizenship. 'Australia still has wonderful opportunities not available elsewhere', he says. He was educated in Scotland at the universities of St Andrews and Edinburgh, and is a Fellow of the Royal Astronomical Society, a member of the Astronomical Society of Australia, and a member of the International Astronomical Union. Fred serves on a number of astronomy-related committees, but he is also a keen advocate for publicly-funded education, and is a Board Member of the Public Education Foundation of NSW.

# The MANIFEST Design Study Proposal

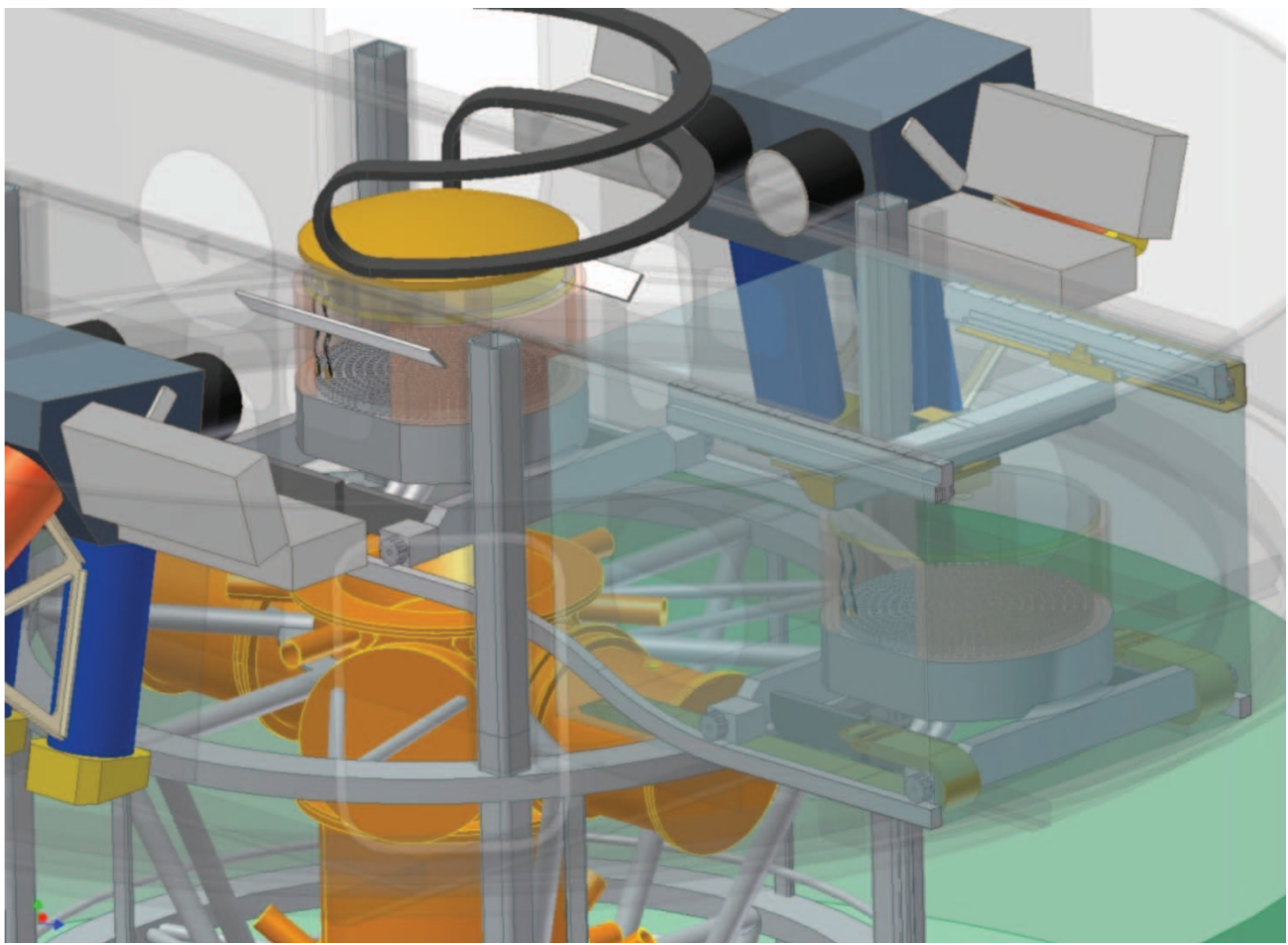

MANIFEST instrument envelope within GMACS on the Giant Magellan Telescope

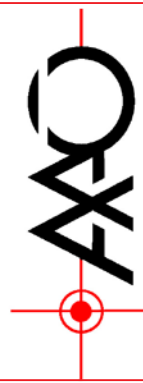